\documentclass[reprint,twocolumn,aps,prd,superscriptaddress]{revtex4-1}

\usepackage{graphicx}         
\usepackage{amssymb}
\usepackage{amsmath}
\usepackage{amsthm}
\usepackage[colorlinks]{hyperref}
\usepackage{color,rotating}
\usepackage{bbm}
\usepackage{array}
\usepackage{pstricks}
\usepackage{pstricks-add}
\usepackage{colonequals}
\usepackage{slashed}
\usepackage{tikz}

\definecolor{darkgreen}{rgb}{0,0.6,0}
\definecolor{gray}{rgb}{.7,.7,.7}

\DeclareMathAlphabet{\EuRoman}{U}{eur}{m}{n}
\SetMathAlphabet{\EuRoman}{bold}{U}{eur}{b}{n}

 \graphicspath{{./figures/}}

\def\di{\displaystyle}

\def\bg{\begin{eqnarray}\begin{array}{rcl}\displaystyle}
\def\eg{\end{array} &\di    &\di   \end{eqnarray}}
\def\bm#1{\begin{eqnarray}\begin{array}{#1}\di}
\def\bmo#1{\begin{eqnarray*}\begin{array}{#1}\di}
\def\bml#1#2{\begin{eqnarray}\begin{array}{#1}\label{#2}\di}
\def\bgo{\begin{eqnarray*}\begin{array}{rcl}\displaystyle}
\def\ego{\end{array} &\di    &\di \nonumber  \end{eqnarray*}}

\def\btensor#1#2{\renew\left#1\begin{array}{#2}\di}
\def\brtensor#1#2#3{\ren#3\left#1\begin{array}{#2}}
\def\botensor#1#2{\renew\left#1\begin{array}{#2}}
\def\etensor#1{\end{array}\right#1}

\def\eq#1{(\ref{#1})}
\def\Eq#1{Eq.~(\ref{#1})}

\def\s0#1#2{\mbox{\small{$ \frac{#1}{#2} $}}}
\def\0#1#2{\frac{#1}{#2}}

\def\s{\sigma}

\def\ra{\rightarrow}

\def\ren#1{\renewcommand{\arraystretch}{#1}}

\def\renew{\renewcommand{\arraystretch}{1}}

\newcommand{\UV}{{\small UV}}
\newcommand{\IR}{{\small IR}}
\newcommand{\FRG}{{\small FRG}}
\newcommand{\RG}{{\small RG}}

\newcommand{\NJL}{{\small NJL}}

\newcommand{\QCD}{{\small QCD}}

\allowdisplaybreaks

\begin{document}

\title{Chiral fermions in asymptotically safe quantum gravity}

%
\author{J. Meibohm}

\affiliation{Institut f\"ur Theoretische Physik, Universit\"at Heidelberg,
Philosophenweg 16, 69120 Heidelberg, Germany}

 \author{J. M. Pawlowski}
 \affiliation{Institut f\"ur Theoretische Physik, Universit\"at Heidelberg,
 Philosophenweg 16, 69120 Heidelberg, Germany}
 \affiliation{ExtreMe Matter Institute EMMI, GSI Helmholtzzentrum f\"ur
 Schwerionenforschung mbH, Planckstr.\ 1, 64291 Darmstadt, Germany}

\begin{abstract}
  We study the consistency of dynamical fermionic matter with the
  asymptotic safety scenario of quantum gravity using the functional
  renormalisation group. Since this scenario suggests strongly coupled
  quantum gravity in the \UV, one expects gravity-induced fermion
  self-interactions at energies of the Planck-scale. These could lead
  to chiral symmetry breaking at very high energies and thus to large
  fermion masses in the \IR. The present analysis which is based on the previous
  works \cite{Christiansen:2015rva, Meibohm:2015twa}, concludes that 
  gravity-induced chiral symmetry breaking at the Planck scale is avoided
  for a general class of \NJL-type models, regardless of the number of
  fermion flavours. This suggests that the phase diagram for these
  models is topologically stable under the influence of gravitational
  interactions.
\end{abstract}
\maketitle

\section{Introduction}
Finding a well-defined theory for quantum gravity is a major challenge
of modern theoretical physics.  The asymptotic safety scenario
\cite{Weinberg:1980gg} is a promising approach towards a solution to
this problem. It relies on the description of quantum gravity in terms
of a local, fundamental quantum field theory of the metric. Within
this scenario the \UV{}- and \IR{} limits of the theory remain well
defined but possibly approach strong-coupling regimes where
perturbation theory is not applicable.

Non-perturbative functional renormalisation group (\FRG) techniques
and their application to quantum gravity
\cite{Reuter:1996cp} allow for detailed studies of
these strong-coupling regimes.  They provide evidence for the
existence, and by now also reveal some of the properties of a
non-trivial \UV{} fixed point of the renormalisation group flow. The latter
controls the \UV{} behaviour of the theory and renders it finite at
arbitrarily high energies.  For pure gravity such a fixed point was
first found in basic Einstein-Hilbert approximations
\cite{Reuter:1996cp,Souma:1999at,Reuter:2001ag} and later confirmed in
more elaborate truncations
\cite{Christiansen:2015rva,Christiansen:2012rx, Christiansen:2014raa,%
Lauscher:2002sq,Codello:2006in,Codello:2007bd,Codello:2008vh,%
Machado:2007ea,Benedetti:2009rx,Eichhorn:2009ah,Manrique:2011jc,Rechenberger:2012pm,%
Donkin:2012ud, Codello:2013fpa,Falls:2013bv,Falls:2014zba, Falls:2014tra,%
Falls:2015qga,Gies:2015tca,Gies:2016con},
for reviews see
\cite{Niedermaier:2006wt,Percacci:2007sz,Litim:2011cp,Reuter:2012id}. First
studies of gravity combined with minimally coupled matter have led to
interesting results and developments 
\cite{Meibohm:2015twa,
Dou:1997fg,Percacci:2002ie,Percacci:2003jz,Folkerts:2011jz,
Dona:2012am,Dona:2013qba,Dona:2014pla,Oda:2015sma, Dona:2015tnf}.
As a key observation, the non-trivial interplay among the fluctuation dynamics
of all involved fields has a crucial impact on the \UV{} behaviour of
the theory.

All theories of quantum gravity have to allow for the inclusion of
dynamical and potentially self-interacting matter
\cite{Eichhorn:2011pc,Eichhorn:2012va, Henz:2013oxa}. In Non-Abelian
gauge theories coupled to matter, the gluon-induced fermion
correlations are responsible for the generation of fermion masses at
low energies.  In these theories, the gauge-coupling becomes large at
low energies and eventually exceeds a critical value. This critical
gauge coupling is responsible for induced strong correlations among
fermions which lead to chiral symmetry breaking and the generation of
fermion masses at low energies. By contrast, gravity becomes strongly
interacting in the ultraviolet within the asymptotic safety
scenario. This raises the question whether there exists a critical
gravitational coupling for which chiral symmetry is broken dynamically
at high energies. Crucially, the generation of fermion masses at high
energies would result in large masses for fermions in the \IR{}. This
is not in agreement with observation, and has to be dealt with in
asymptotically safe theories of gravity with matter. The question arises whether
there are mechanisms at work that prevent a theory of fermions and
gravity from being driven to criticality. One scenario is, that the
critical coupling is never reached for general initial conditions.
In a less restrictive scenario, the system could in principle reach
criticality but only for an unphysical set of parameters. 

In a similar study \cite{Eichhorn:2011pc}, the authors find no
indications for chiral symmetry breaking in a combined setup of a flat
expansion and a background field approach including only a
non-dynamical fermion anomalous dimension. In the present work, we
reconsider the question of gravity-induced chiral symmetry at energies
of the order of the Planck scale on a more general basis in the
self-consistent vertex expansion scheme put forward in
\cite{Christiansen:2012rx,Christiansen:2014raa,%
  Christiansen:2015rva,Meibohm:2015twa}. The results are, to a large
degree, regularisation scheme independent and also include dynamical
anomalous dimensions for all field species. We show that the phase
diagram of \NJL-type interacting fermion theories is topologically
stable under the influence of asymptotically safe quantum gravity. Our
study implies that metric gravity and the asymptotic safety scenario
stay consistent under the inclusion of an arbitrary number of
fermions with a point-like 4-fermion interaction.

\section{Quantum Fluctuations in gravity with fermionic matter}
\begin{figure}[t]
\centering
  \includegraphics[width=8cm]{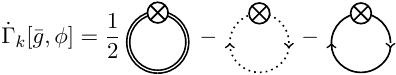}
  \caption{Flow equation for the scale dependent effective action
    $\Gamma_k$ in diagrammatic representation. The double, dotted, and
    solid lines correspond to the graviton, ghost, and fermion
    propagators, respectively. The crossed circles denote the
    respective regulator insertions.}
  \label{f:wetterich}
\end{figure}

The quantum effective action is the quantum analogue of the classical
action and encodes all quantum fluctuations of the theory. In the
present case of gravity coupled to fermionic matter we have the
effective action $\Gamma[\bar g,\phi]$ with vertices $\Gamma^{(n)}$,
the amputated $n$-point correlation functions.  The effective action
depends on a background metric $\bar g$ and the fluctuation field
$\phi$ which comprises all fluctuating gravity and matter fields. The
full metric field $g_{\mu\nu}$ is split linearly into background and
fluctuating fields, $\bar g_{\mu\nu}$ and $h_{\mu\nu}$, respectively.
For the present theory of gravity and fermions, the complete set of
fluctuating fields is given by
\begin{align}\label{e:fluct}
\phi=(h,\bar c,c, \bar\psi,\psi)\,.
\end{align}
Here, $(\bar c,c)$ denote the (anti-)ghosts and $(\bar \psi,\psi)$ are
the (anti-)fermion fields. By adding a scale dependent \IR{}-regulator
term $R_k[\bar g,\phi]$ to the classical action we obtain the
scale-dependent effective action $\Gamma_k[\bar g,\phi]$. It gives an
effective description of the physics at scale $k$ in the spirit of the
Wilsonian renormalisation group. The flow of $\Gamma_k[\bar g,\phi]$
is governed by the Wetterich equation \cite{Wetterich:1992yh}, applied
to gravity \cite{Reuter:1996cp}. For the given field content \eqref{e:fluct}
it reads
\begin{align}\label{eq:flow}
  \begin{split}
    \dot \Gamma_k[\bar g,\phi]& = \frac{1}{2} \text{Tr}\left[ \frac{
        1}{\Gamma_k^{(2)}+R_{k}}\dot R_{k} \right]_{hh}\\
    - &\text{Tr}\left[ \frac{1}{\Gamma_k^{(2)}+R_{k}}\dot R_{k}
    \right]_{\bar c c} \hspace*{-5pt} - \text{Tr}\left[
      \frac{1}{\Gamma_k^{(2)}+R_{k}}\dot R_{k} \right]_{\bar \psi
      \psi} \hspace*{-5pt}\,,
  \end{split}
\end{align}
where $\text{Tr}$ denotes the summation over discrete and integration
of continuous variables. We abbreviate by a dot derivatives with
respect to $t=\log(k/k_0)$, where $k_0$ is some arbitrary reference
scale.  \autoref{f:wetterich} depicts equation \eqref{eq:flow} in
terms of diagrams.
From now on, we will drop the regulator insertions in the diagrams as
well as the arrows for Grassmann-valued fields for convenience.

\Eq{eq:flow} cannot be solved for $\Gamma_k[\bar g,\phi]$ in full
generality. Therefore, the effective action is typically truncated to
a finite set of functionals $\mathcal{O}_i$ to wit
\begin{align}
  \Gamma_k[\bar g,\phi]=\sum_{i=0}^N \bar\alpha_i(k)
  \mathcal{O}_i[\bar g,\phi]\,.
\end{align}
With help of this ansatz the flow equation \eqref{eq:flow}
provides a finite dimensional coupled system of flow equations for the
$k$-dependent dimensionful couplings $\bar\alpha_i(k)$ with mass
dimension $d_i$. Dimensionless couplings $\alpha_i$ are introduced by
dividing $\bar \alpha_i$ with $k^{d_i}$. The flow equations of the
dimensionful couplings $\bar \alpha_i(k)$ of mass dimension $d_i$ are
related to the flow of their dimensionless counterparts $\alpha_i(k)$
by
\begin{align}
 k^{-d_i} \dot{\bar \alpha}_i(k) =  d_i \alpha_i + \dot {\alpha}_i\,.
\end{align}
The truncation applied in this work is constructed along the same
lines as in \cite{Christiansen:2014raa,Christiansen:2015rva,
  Meibohm:2015twa}. Thus, $\Gamma_k[\bar g,\phi]$ is given in terms of
a vertex expansion in the fluctuating fields $\phi$ about the
expansion point $\phi=0$ with the structure  
\begin{align}\label{e:vexp}
  \Gamma_k[\bar g, \phi] =
  \sum_{n=0}^{\infty}\frac{1}{n!}\Gamma^{(n)}_k[\bar
  g,0]\circ\phi^{n}\,, 
\end{align}
where the circle indicates the integration of $\phi(x_1)\cdots
\phi(x_n)$ with $\Gamma^{(n)}(x_1,...,x_n)$. As can be seen from
\eq{eq:flow}, the two-point functions plays a pivotal r$\hat{\rm o}$le
in the current approach. We parameterise 
\begin{align}\label{eq:2point} 
	\Gamma_k^{(\phi\phi)}(p^2)= Z_\phi(p^2) \mathcal{T}^{(2)}\,,
\end{align}
where $ \mathcal{T}^{(2)}$ carries the tensor structure of the
two-point function, see \eq{eq:Tn}, and $Z_\phi$ are the
momentum-dependent wave function renormalisations.  Similarly, the
vertices $\Gamma_k^{(n)}$ are 
parameterised as 
\begin{align}\nonumber 
  \Gamma^{(n)}_k[\bar g,0]=&
  \left(\prod_{i=1}^{n_h}\sqrt{Z_h(p_{h,i}^2)}\right)\,\left(\prod_{i=1}^{n_c}
    \sqrt{Z_c(p_{c,i}^2)}\right)\\[2ex]  & \times
  \left(\prod_{i=1}^{n_\psi}\sqrt{Z_\psi(p_{\psi,i}^2)}\right)\,
  G_{\vec n}^{\frac12(n-2)}\mathcal{T}^{(n)}\,,
\label{eq:vert} \end{align}
where $\vec n=(n_h,n_c,n_\psi)$ and $n=n_h+n_c+n_\psi$. Here, $n_h$,
$n_c$ and $n_\psi$ denote the number of graviton, ghost and fermion
legs, respectively. The wave function renormalisations in \eq{eq:vert}
carry the RG-running of the vertices. Indeed, for the two-point
functions \eq{eq:vert} boils down to \eq{eq:2point}. For $n\neq 2$ the
$G_{\vec n}$ are renormalisation group invariant couplings.  For example, for
the pure gravitational vertices, $n=n_h+n_c$ the $G_{\vec n}$ are
momentum dependent Newton's constants $G_{n}(\mathbf{p})$. Here,
$\mathbf{p}$ is the vector of all field momenta. The vertices that 
implement the minimal coupling between gravity and fermions are consequently
associated with $G_{n_h,0,2}(\mathbf{p})$. In the same way as in
\cite{Christiansen:2015rva, Meibohm:2015twa}, however, we approximate
here all $G_n$ as one, momentum-independent coupling,
$G_{n_h,n_c,0}(\mathbf{p}) = G_{n_h,0,2}(\mathbf{p})\equiv G_{3,0,0} =: G$. In the four-fermion
case we have $G_{n,0,4}=\lambda_\psi$.

The tensor structures $\mathcal{T}^{(n)}$ are given by the
$n^\text{th}$ variation of the classical action $S$ to be specified
below. The couplings and wave function renormalisations appearing in
$S$ are replaced by their $k$-dependent counterparts. Bearing this in
mind we write the $\mathcal{T}^{(n)}$ as
\begin{align}
  \mathcal{T}^{(n)}=S^{(n)}\left(\mathbf{p},\Lambda\rightarrow\Lambda_{n}(k),G_N\rightarrow1,
    \bar\lambda_\psi\rightarrow1\right)\,,
\label{eq:Tn}\end{align}
where $\Lambda$ is known as the dimensionful cosmological constant and
$G_N$ is the classical gravitational constant.
We set the classical couplings $G_N$ and $\bar\lambda_\psi$ to one since they
are replaced by the running couplings $G_{\vec n}$ in the vertex parametrisation
\eqref{eq:vert}. The $k$-dependence of all running quantities will be dropped in the
following and is understood implicitly. We extract the anomalous
dimensions for the fields as well as the flows for all couplings from
the flows of the $n$-point vertices. To this end, we substitute our
truncation ansatz \eqref{e:vexp} into the $n^\text{th}$ variation of
\eqref{eq:flow}.  Evaluating all quantities in a flat metric
background $\bar g_{\mu\nu}=\delta_{\mu\nu}$ and at vanishing field
momenta $\mathbf{p}=0$ we compare coefficients to obtain flow
equations for the dimensionful couplings
$(G,\Lambda_n,\bar\lambda_\psi)$. In a similar manner, we obtain the
flows for the wavefunction renormalisations $Z_\phi$ which, however,
appear in the flow only in terms of the anomalous dimensions
\begin{align}
	\eta_\phi (p^2)	=	-\partial_t \ln (Z_\phi (p^2))\,.
\end{align}
We achieve the latter simplification by introducing regulators
$R^\phi_k(p^2)$ which depend on the momentum dependent part of the
2-point functions and, consequently, on the wavefunction
renormalisation of the respective fields according to
\begin{align}\label{e:regclass}
  R^\phi_k(p^2) = \Gamma_k^{(\phi\phi)}(p^2)\bigg|_{\Lambda_2=0}
  r^\phi_k(p^2)\,,
\end{align}
where $r^\phi_k$ denotes the regulator shape function. Details on this
general class of regulators can be found \cite{Meibohm:2015twa}.

Consequently, the complete system is governed by the flow of the three
couplings given above and a coupled system of equations for the
anomalous dimensions $(\eta_h,\eta_c,\eta_\psi)$.
\begin{figure}
\centering
  \includegraphics[width=4.5cm]{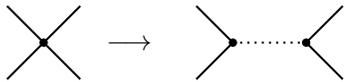}
\caption{Transition of a point-like 4-fermion coupling to
  scalar-field-mediated interaction.}
\label{f:hub_strat}
\end{figure}
\subsection{Classical Action}
The classical action $S$ is given by the sum
of the gauge-fixed Einstein-Hilbert action $S_\text{EH}$ which governs
the dynamics of classical gravity and the fermionic action
$S_\text{ferm}$. The latter contains the fermion kinetic term and
describes the classical 4-fermion interactions. Hence, we write $S$ as
\begin{align}
  S=S_\text{EH}+S_\text{ferm}\,.
\end{align}
The well-known $S_\text{EH}$ reads
\begin{align}\label{e:EH}
  S_{\text{EH}}=\frac1{16\pi G_N}\int\left(2\Lambda-R\right)\sqrt{g}\,d^4x
  +S_\text{gf}+S_\text{gh}\,,
\end{align}
where we $G_N$ denotes the classical Newton
coupling. Furthermore, in
\eqref{e:EH} $S_\text{gh}$ and $S_\text{gf}$ denote the ghost action
and the gauge fixing term, respectively. The latter are determined by
the gauge condition $F_\mu$. We apply a De-Donder-type linear gauge in
the Landau-limit of vanishing gauge parameter. More precisely, the
gauge-fixing condition is given by
\begin{align}
  F_\mu = \overline\nabla^\nu h_{\mu\nu} - \frac{1+\beta}{4}
  \overline\nabla_\mu {h^\nu}_\nu \, ,
\end{align}
with $\beta=1$ in this work. The Landau-limit is particularly
convenient since it provides a sharp implementation of the gauge
fixing. This assures that the corresponding gauge-fixing parameter is
at a fixed point of the renormalisation group flow
\cite{Litim:1998qi}. The fermionic action $S_\text{ferm}$ reads
\begin{align}
  S_{\text{ferm}} =\int
  \Bigl(\bar\psi^i\slashed{\nabla}\psi^i +
    V(\psi,\bar\psi)\Bigr)\sqrt{g}\,d^4x\,,
\end{align}
where $V(\phi,\bar\phi)$ denotes a fermionic potential with chiral symmetry. Thus, $V(\phi,\bar\phi)$ is
invariant under the axial rotation
\begin{align}\label{eq:axial}
  \psi\to e^{i \alpha\gamma_5}\,\psi\qquad {\rm and} \qquad \bar
  \psi\to \bar \psi\, e^{i \alpha\gamma_5 }\,.
\end{align}
Invariance under \eq{eq:axial} also excludes a fermionic mass
term. Then, the fermion kinetic term is governed by the spin-covariant
derivative $\slashed \nabla$ given by
\begin{align}\label{e:spin_base}
  \slashed \nabla =g_{\mu\nu}\gamma(x)^\mu( \partial^\nu +
  \Gamma(x)^{\nu} )\,.
\end{align}
Here, $\Gamma^\mu$ denotes the matrix-valued spin connection. The
kinetic term is invariant under \eq{eq:axial} and, therefore, is chirally
symmetric. The
operator $\slashed \nabla$ is defined using the spin-base invariant
formalism introduced in \cite{Weldon:2000fr,Gies:2013noa,
  Lippoldt:2015cea} which avoids the use of a vielbein.  In particular, it allows to
circumvent vielbein-related gauge ambiguities.  The latter
formulation relies on spacetime-dependent spin connection and gamma
matrices, which is made explicit in \eqref{e:spin_base}. Note, that we
will drop the spacetime-dependence
from the notation in the following for convenience.
\begin{figure}
\centering
  \includegraphics[width=8.5cm]{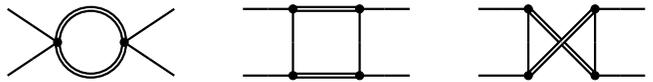}
\caption{Gravity-induced contributions to the flow of the 4-fermion
  coupling. The diagrams generate a non-zero 4-fermion
  self-interaction dynamically, even though the corresponding
  4-fermion couplings $\lambda_i$ might be zero at some energy scale.}
\label{fig:ind_coup}
\end{figure}
\subsection{\NJL-model with one Fermion}
As a simple, Fierz-complete
model for interacting fermions, we consider the \NJL{}-model with
one fermion flavour $N_f=1$. Its potential is given by
\begin{align}\label{e:NJLpot}
  V(\psi,\bar\psi)
  =\frac12\bar \lambda_v \left[(\bar\psi\gamma_\mu\psi)^2\right]
  -\frac12\bar\lambda_\sigma\left[(\bar\psi \psi)^2 -
    (\bar\psi\gamma_5\psi)^2\right]\,, 
\end{align}
which is invariant under the axial rotations, \eq{eq:axial}. While the first
term in \eqref{e:NJLpot} is trivially invariant due to $\{\gamma_5,\gamma_\mu\}=0$,
the scalar and pseudoscalar terms rotate into each other under
\eq{eq:axial}. The first and second terms in \eqref{e:NJLpot}
represent vector and scalar-pseudoscalar interactions,
respectively. We resolve the 4-fermion coupling $G_{0,0,4}$
individually for the two latter interaction channels as
$\bar\lambda_v$ and $\bar\lambda_\sigma$.  In the
present work we consider momentum-independent
$\bar\lambda_v$ and $\bar\lambda_\sigma$. This assures that the
4-fermion interactions do not couple back to the minimally-coupled sub-system of fermions
and gravity which comprises the sub-system of
flows $(\partial_t G,\partial_t
\Lambda_i,\eta_h,\eta_c,\eta_\psi)$. This crucial feature of the present
truncation will be discussed in more detail below.

\begin{figure*}
  \centering
  {\raisebox{1.75pt}{\includegraphics[width=.9
\columnwidth]{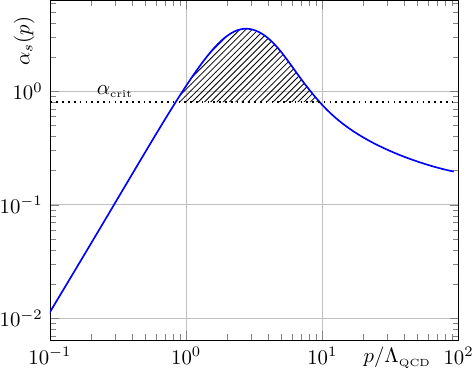}}\hspace{1.5cm}
\includegraphics[width=.9\columnwidth]{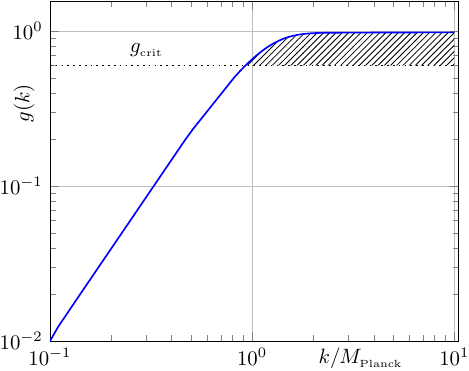}}
\caption{Left: Running of the strong coupling $\alpha_s$ as a function
  of the energy scale $p/\Lambda_\text{QCD}$ as calculated in
  \cite{Mitter:2014wpa}. In order for chiral symmetry breaking to take
  place, $\alpha_s$ must exceed the critical coupling
  $\alpha_{\text{crit}}$ in a sufficiently large interval of
  $p/\Lambda_\text{QCD}$ (shaded region). Right: Running of the
  gravitational coupling $g$ calculated from the analytical flow
  equations given in \cite{Christiansen:2015rva}. The shaded region
  represents the interval, where chiral symmetry breaking could take
  place provided there exists a critical value for $g(k)$,
  $g_\text{crit}$, which lies below the threshold of $g(k)$.}
\label{fig:al_k}
\end{figure*}
\section{Chiral Fermions and Asymptotic Safety}
\subsection{Chiral Symmetry Breaking}
In this work, we study
gravity-induced chiral symmetry at Planck-scale energies. The
spontaneous breaking of chiral symmetry occurs if a non-zero
expectation value $\langle \bar\psi\psi\rangle$ is dynamically
generated. This necessitates a divergence of the dimensionless
4-fermion coupling $\lambda$ according to
\begin{align}\label{e:chisb}
	\frac{1}{\lambda}\rightarrow 0\,.
\end{align}
The origin of this mechanism is most apparent in a partially bosonised
description, where the point-like 4-fermion interaction is replaced by
the interaction with a scalar field $\varphi$ by means of a Hubbard-Stratonovich
transformation. In
\autoref{f:hub_strat}, this procedure is depicted in terms of diagrams,
where the propagation of $\varphi$ is represented by a dashed line.  As a result of the transformation, the
momentum dependence of the 4-fermion vertex is modelled by the
mediation of $\varphi$.  The scalar mass $m_\varphi$ is now related to
4-fermion coupling by $\frac1{\lambda}\sim{m_\varphi^2}$. Hence, in
this light, equation \eqref{e:chisb} corresponds to a change of sign
for the mass term $m^2_\varphi$. This is interpreted as the
acquirement of a Mexican-hat shape for the scalar potential which
leads to a non-zero vacuum expectation value for $\varphi$ and thus
for $\bar\psi\psi$.

In the following, however, we leave the bosonised formulation aside
and study directly the flow of the dimensionless 4-fermion
couplings $\lambda_i$ with $i=v,\sigma$ of the potential
\eqref{e:NJLpot}. This allows us to investigate whether chiral symmetry
breaking according to \eqref{e:chisb} is in fact realised in the
present model, see e.g. \cite{Gies:2005as,Braun:2011pp}. At one-loop level, the 4-fermion interaction is
generated by the interaction with gravity via the diagrams given in
\autoref{fig:ind_coup}, and hence is proportional to the Newton
coupling squared, $G^2$. 

This structure is reminiscent of that leading to chiral-symmetry
breaking in \QCD{}, where it generates the large low-energy
constituent quark masses. In the latter case, the gauge coupling
$\alpha_s$ acquires a critical value in the infrared.  The left panel
in \autoref{fig:al_k} depicts the running of the strong coupling
$\alpha_s$ as a function of the energy scale $p/\Lambda_\text{QCD}$ as
calculated in \cite{Mitter:2014wpa}.

The analysis in \cite{Mitter:2014wpa} favours a scenario in which
$\alpha_s$ becomes critical $\alpha_s(p^2)>\alpha_\text{crit}$ when
running towards the \IR, but then becomes subcritical again at lower
momenta. It is argued that $\alpha_s$ must stay in the critical regime
(shaded region) for a momentum scale window that is large enough for
chiral symmetry breaking to take place. In the case of gravity, the
dimensionless gravitational coupling $g=G k^2$ becomes stronger with
increasing \RG-scale. \autoref{fig:al_k} shows $g$ as a function of
$k/M_\text{Planck}$, calculated from the analytical equations in
\cite{Christiansen:2015rva} without the inclusion of matter.
If a critical coupling $g_\text{crit}$ exists, it is most certainly
reached only at $k\approx M_\text{Planck}$. Hence, in contrast to the
case in Yang-Mills theories, the existence of a critical coupling for
gravity would lead to the generation of fermion masses at energies of
the order of the Planck scale. This however, leads to fermion masses
of the order of the Planck scale in the \IR, which are not observed in
nature.

\subsection{Asymptotic Safety}
It is well-known that the \NJL{}-model
without gravity interactions exhibits a (Gaussian) \IR{}-stable fixed
point of the renormalisation group flow. Furthermore, theories of the
latter kind have a well defined \UV{}-limit due to the existence of
several (non-Gaussian) fixed points with \UV{}-stable directions. Thus
there exist well defined limits both for the \IR, $k\to0$, and the
\UV, $k\to\infty$.

On the other hand, there is strong evidence that quantum gravity
defined in terms of a quantum field theory of the metric, is
asymptotically safe. Therefore, the latter has a well defined,
although strongly coupled \UV{}-limit controlled by a \UV{}-attractive
fixed point. The usual Gaussian fixed point always features one
\IR{}-attractive direction. Careful studies of the \IR{}-behavior of
quantum gravity even suggest the existence of more \IR{}-fixed points,
apart from the Gaussian fixed point \cite{Christiansen:2014raa}. The
purpose of this work is now to analyse the \UV{}- and \IR{}-behaviour
of the - minimally coupled - combined theory of interacting fermions
and quantum gravity. The minimal coupling connects both theories via
the kinetic term in the fermionic action $S_\text{ferm}$, which is a
function of the fluctuating graviton field $h$.

\begin{figure}
\centering
  \includegraphics[width=8cm]{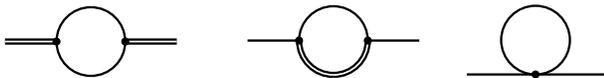}
	\caption{Loop-contributions to the flow of the 2-point
          functions which involve interactions of the fermion
          field.}\label{f:anomdiag}
\end{figure}

First, we discuss in which way the two theories interfere due to the
minimal coupling. To this end, we neglect the fermion interaction due
to $V(\psi,\bar\psi)$. In a theory with $V(\psi,\bar\psi)=0$ the
flows for the graviton and fermion 2-point functions receive mutual
loop contributions due to their minimal coupling. The first two diagrams in \autoref{f:anomdiag}
depict possible contributions to the latter flows.
The resulting interaction of free fermions with gravity is known to
alter the \UV{}-behaviour of the complete theory considerably
\cite{Codello:2008vh,Dona:2013qba,Meibohm:2015twa}. However, in all
approaches the inclusion of one fermion is no threat to the existence
of the \UV{}-fixed point. Hence, the combined theory of free fermions
minimally coupled to gravity remains well defined.

In the remainder of this work we consider interacting fermions thus
$V(\psi,\bar \psi)\neq0$. Note that the 4-fermion interaction along with all
$2n$-fermions are generated dynamically due to the interaction with
gravity in the loops. Hence, truncations which disregard the latter
interactions cannot capture this part of the flow. The fermion
potential \eqref{e:NJLpot} discussed in this work includes a Fierz-complete
basis of momentum independent 4-fermion interactions, thereby
neglecting the higher $2n$-fermion vertices.

As mentioned above, non-vanishing 4-fermion interactions
$V(\psi,\bar\psi)\neq0$, do not alter the flow of the minimally-coupled sub-system
of fermions and gravity.  This can be understood by considering the possible
contributions to the flow of the $n$-point functions that play a r\^{o}le
here. The 4-fermion interaction enters the flow of the lower order
$n$-point functions only via the third diagram in
\autoref{f:anomdiag}.  The latter diagram can only
contribute to the fermion anomalous dimension $\eta_\psi$. It does not, however,
since the 4-fermion interaction is momentum independent and no external momentum
runs in the loop.  The resulting decoupling of the flow of the 4-fermion
interaction from the rest of the system manifests itself in the fact,
that the minimally-coupled sub-system of fermions and gravity, $(\partial_t G,\partial_t \Lambda_i,\eta_h,
\eta_c, \eta_\psi)$, does not depend on the 4-fermion couplings
$\bar\lambda_\sigma$ and $\bar\lambda_v$.

On the other hand, we will see that the flow of the 4-fermion interaction is
influenced by the interaction with gravity. Hence, the task is to
study which impact the gravity interactions have on the \UV{}- and \IR{}-behaviour of the
4-fermion theory and, thus, the existence of fixed points of the
beta-functions
$(\beta_{\lambda_\sigma},\beta_{\lambda_v})=(\partial_t{\lambda_\sigma},
\partial_t{\lambda_v})$ and their stability. The latter are properties
of the phase diagram and closely linked to the onset of chiral
symmetry breaking during the flow from the \UV{} towards the \IR.
\section{Flow Equations}
\begin{figure}
\centering
  \includegraphics[width=8.5cm]{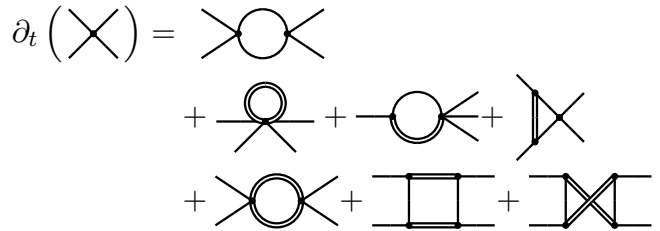}
  \caption{The flow of the 4-fermion coupling given in terms of
  diagrams. The first, second and third line constitute contributions
  proportional to $g^0\lambda_i^2$, $g^1\lambda_i^1$ and
  $g^2\lambda_i^0$, respectively.}
\label{fig:diagflow}
\end{figure}

In terms of diagrams, the flow of the 4-fermion interaction is given
in \autoref{fig:diagflow} where we have ordered the diagrams according
to the their powers in $G$ and $\bar\lambda_i$.
The first line of \autoref{fig:diagflow} governs the contributions of
the \NJL{}-model without gravity. This part is quadratic in the
4-fermion couplings $\bar \lambda_i$ and originates from only one
diagram.  In the second line we have the contributions linear in $G$
and $\bar\lambda_i$, several diagrams contribute here.  In line three
the contributions quadratic in $G$ and independent from
$\bar\lambda_i$ are given. Since the last two diagrams cancel for
vanishing fermionic masses, the contribution quadratic in $G$ is also
governed by only one diagram. Note that another triangle-type diagram
with two graviton propagators does in principle contribute to the
third line of \autoref{fig:diagflow}. However, this vanishes in the
here considered Landau limit $\alpha\rightarrow0$.

From the diagrammatic flow (\autoref{fig:diagflow}), we extract the
flow equations for the couplings $\bar \lambda_\sigma$ and $\bar
\lambda_v$. In the following, we will express all equations in terms
of dimensionless quantities. In particular, the dimensionsless Newton
coupling, the dimensionless graviton-mass parameter and the
dimensionless 4-fermion coupling are written as $g=G k^2$,
$\mu=-2\Lambda_2/k^2$ and $\lambda_i=\bar\lambda_i k^2$, respectively. Using
the latter, we arrive at the $\beta$-functions for $\lambda_\sigma$
and $\lambda_v$. They read
\begin{align}\label{e:lamsys}
\begin{split}
  \beta_{\lambda_\sigma} \hspace*{-3pt}&= 2\mathfrak{h}\lambda
  _\sigma^2+2(1+
  \eta_\psi(0)+\mathfrak{f}+4\mathfrak{h}\lambda_v)\lambda_\sigma +
  6\mathfrak{h}\lambda_v^2+\mathfrak{g}\,,\\
  \beta_{\lambda_v} \hspace*{-3pt}&= \mathfrak{h}\lambda
  _v^2+2(1+\eta_\psi(0)+\mathfrak{f}+\mathfrak{h}\lambda_\sigma)\lambda_v
  +\mathfrak{h}\lambda_\sigma^2-\frac12\mathfrak{g}\,,
\end{split}
\end{align}
which agrees with \cite{Braun:2011pp} if we neglect the gravity contributions.
Since we evaluate the flow of the 4-fermion vertices at vanishing
external momenta, $\eta_\psi(0)$ appears explicitly in the
equations. Due to the structure of the 4-fermion interaction we were
able to simplify the flow equations considerably, by identifying
different terms corresponding to structurally identical diagrams. The
functions $\mathfrak{h}$, $\mathfrak{f}$ and $\mathfrak{g}$ correspond
to the diagrams depicted in the first, second and third line of
\autoref{fig:diagflow}, respectively.  The quantities $\mathfrak{f}$
and $\mathfrak{g}$ still depend on the gravitational coupling $g$ and
can be written as $\mathfrak{f}(g)=g \mathfrak{f}(1)$ and
$\mathfrak{g}(g)=g^2\mathfrak{g}(1)$. As a result, $\mathfrak{f}$ and
$\mathfrak{g}$ vanish in the limit $g\to0$. The same is true for the
fermion anomalous dimension, $\eta_\psi$.  Since $\eta_\psi$,
$\mathfrak{f}$ and $\mathfrak{g}$ are the only quantities in our
analysis which depend on $g$ and $\mu$, these three objects parametrise
completely the interaction of the 4-fermion system with the minimally-coupled gravity-fermion sub-system.

The functions $\mathfrak{g}$ and $\mathfrak{h}$, correspond to the
first and third line in \autoref{fig:diagflow}, respectively. Thus,
the latter both originate from only one diagram. Since the signs of
single diagrams do not change under the change of regulator,
$\mathfrak{g}$ and $\mathfrak{h}$ also have fixed signs according to
\begin{align}\label{e:fsigns}
  0>\mathfrak{g}	&=	-48\pi g^2\frac{v_3}{(2\pi)^4}
\int_0^\infty\frac{\dot r_k^{(h)}(q)-\eta_h(q^2) r_k^{(h)}(q)}{(q^2(1
+r_k^{(h)}(q))+\mu)^2}q^3\text{d}q\,,	\notag\\
0>\mathfrak{h} &= -2 \frac{v_3}{(2\pi)^4}\int_0^\infty \frac{\dot
  r_k^{(\psi)}(q)-\eta_\psi(q^2) r_k^{(\psi)}(q)}{(q^2(1
  +r_k^{(\psi)}(q)))^2}q^3\text{d}q\,.
\end{align}
Here, $v_3=2\pi^2$ is the volume of $\mathbb{S}_3$. By contrast,
$\mathfrak{f}$ originates from the three diagrams in the second line
of \autoref{fig:diagflow}. This means that $\mathfrak{f}$ does in
general not have a fixed sign.
As we show below, the fixed signs of $\mathfrak{g}$ and $\mathfrak{h}$
allow to make regulator independent statements about the fixed points
of $(\beta_{\lambda_\sigma},\beta_{\lambda_v})$.

\begin{figure}
  \centering
 \includegraphics[width=8.5cm]{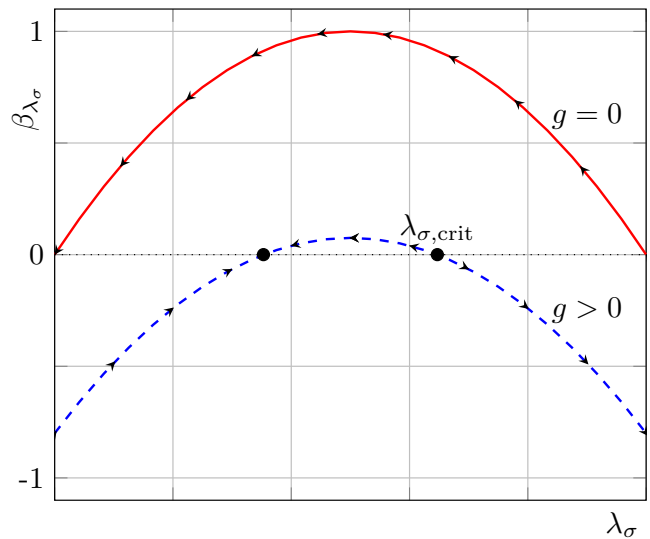}
 \caption{Schematic shape of $\beta_{\lambda_\sigma}$ as a function of
   $\lambda_\sigma$ for fixed $\lambda_v$, where the arrows denote the
   direction of the flow towards the \IR{}. For $g^*=0$, $\lambda_v=0$
   the system exhibits two fixed points. The first one is the IR
   attractive (Gaussian) fixed point, the second one is IR-repulsive.
   Non-zero $g^*$ or $\lambda_v$ lead to up- or down-shift
   as well as a deformation of $\beta_{\lambda_\sigma}$.  }\label{f:lam_parab}
\end{figure}

In order to understand the general behaviour of the flows of
$\lambda_\sigma$ and $\lambda_v$, we consider first the flow of
scalar-pseudoscalar coupling $\lambda_\sigma$ for fixed
$\lambda_v$. For fixed $\lambda_v$ the system of flow equations
\eqref{e:lamsys} reduces to only one equation for
$\beta_{\lambda_\sigma}$. Since the latter is a quadratic function of
$\lambda_\sigma$, it has a generic parabola-shape which is depicted
schematically in \autoref{f:lam_parab}.
The red solid curve represents $\beta_{\lambda_\sigma}$ for vanishing
$\lambda_v=0$ and vanishing dimensionless gravitational coupling $g=G
k^2=0$. The dashed blue curve corresponds to the case of non-vanishing
$\lambda_v$ and $g$. The $\beta$-function for the reduced system with
fixed $\lambda_v$ admits two fixed points (black dots), the arrows
represent the corresponding flow in the \IR{}-direction. As the arrows
suggest, the left fixed point is \IR{} attractive, whereas the right
one is \IR{} repulsive. In the present case ($\lambda_v=0$ and $g=0$)
the left fixed point is trivial (Gaussian). During the flow towards
the \IR, the 4-fermion coupling $\lambda_\sigma$ diverges if its value
in the \UV{} exceeds the critical value
$\lambda_{\sigma,\text{crit}}$. The latter is given by the value of
the repulsive (right) fixed point. In particular, non-zero values of
$g$, as given e.g. in \autoref{fig:al_k}, alter
$\lambda_{i,\text{crit}}$ in the ultraviolet. Thus, the former result
in an up- and downshift of the solid red curve in
\autoref{f:lam_parab}, dependent on the sign of the
$\lambda$-independent contributions to $\beta_{\lambda_\sigma}$. In
\eqref{e:lamsys} the $\lambda$-independent contributions are given by
$\mathfrak{g}$ and we have argued above that $\mathfrak{g}$ is
negative for all regulators. This suggests a $g^2$-dependent downshift
of $\beta_{\lambda_\sigma}$, represented by the blue dashed curve in
\autoref{f:lam_parab}. The formerly Gaussian fixed point is driven to
non-zero values which results in a non-zero flow for the $\lambda_i$
at $(\lambda_\sigma,\lambda_v)=(0,0)$. Hence, even if the 4-fermion
coupling is zero at some (e.g. cutoff) scale it is always dynamically
generated by the flow.

\begin{figure}
  \centering
 \includegraphics[width=8.5cm]{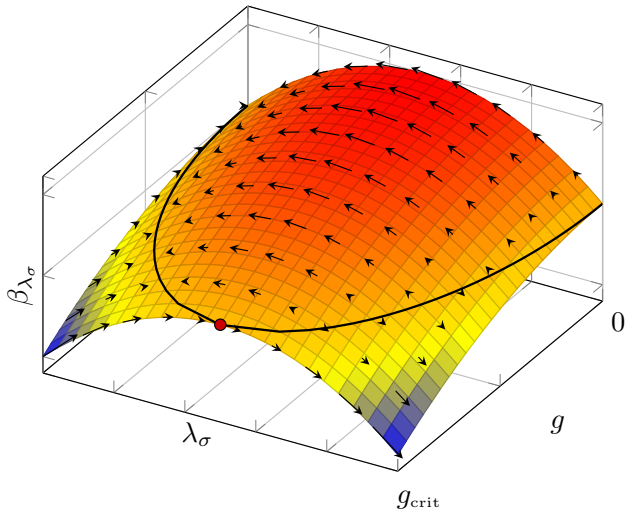}
 \caption{Schematic shape of $\beta_{\lambda_\sigma}$ as a function of
   $\lambda_\sigma$ and $g$ for fixed $\lambda_v$. The black line
   represents the two possible fixed points of the system. For
   large-enough $g$ the two fixed point lines annihilate (red
   dot).}\label{f:lam_parab3d}
\end{figure}

Due to $\mathfrak{g}\neq0$ the two possible fixed points of the
reduced system approach each other and annihilate, if $g$ reaches a
critical value $g_\text{crit}$. This scenario is in accordance with
the assumption that gravity favours strong correlations between
fermions and, thus, chiral symmetry breaking. In
\autoref{f:lam_parab3d} the $\beta$-function for $\lambda_\sigma$ is
plotted schematically as a function of $\lambda_\sigma$ and $g$.
The black line represents the fixed points as a function of $g$. In
the present na\"{i}ve picture of one 4-fermion coupling the two fixed
points annihilate (red dot) if $g$ and, hence, $\mathfrak{g}$ are
large enough. In this case $\beta_{\lambda_\sigma}$ becomes negative
for all values of $\lambda_\sigma$ which is a strong indication for a
divergence of $\lambda_\sigma$. In this scenario, the critical value
$g_\text{crit}$ is given by the strength of the gravitational
coupling $g$ which is necessary in order to result in an annihilation of
the fixed points of $\beta_{\lambda_\sigma}$. If $g$ stays below
$g_\text{crit}$ during the whole \RG-flow, the flow of
$\lambda_\sigma$ is equipped with well defined \IR{} and \UV{} limits,
$k\ra0$ and $k\ra\infty$, respectively. In this case the gravitational
coupling is not strong enough to drive the 4-fermion interactions to
criticality and induce chiral symmetry breaking.

Generally, the requirement that the 4-fermion coupling $\lambda$
exceeds its critical value is a necessary albeit not a sufficient
criterion for chiral symmetry breaking. Since the gravitational
coupling $g$ is a function of the scale parameter $k$, the critical
coupling $\lambda_{\text{crit}}$ changes with the flow as well. This
means that it is in principle possible that a quickly varying $g(k)$
alters $\lambda_\text{crit}$ such that a 4-fermion coupling which had
already exceeded its critical value at some scale, say $k_1$, becomes subcritical again
at some lower scale $k_2$ with $k_2<k_1$. This
behaviour is observed for e.g. Yang-Mills theories as discussed in the previous section (see \autoref{fig:al_k}). The sufficient
criterion therefore states that the gravitational coupling $g$ stays
above its critical value $g_\text{crit}$, which defines
$\lambda_\text{crit}$, for a sufficient amount of \RG-time $t_\text{crit}=\log(k_1/k_2)$. Within
this $t$-interval, $\lambda$ grows rapidly. Thus, if $t_\text{crit}$ is large
enough, $\lambda$ grows so large that it exceeds $\lambda_\text{crit}$ at all following
\RG-times.

For simplicity
however, we will stick for now with the necessary condition of $\lambda>\lambda_\text{crit}$ given above,
not taking into account the time spent in the critical $\lambda$-regime. Clearly,
if the necessary condition for the onset chiral symmetry breaking is
not fulfilled, there is no need to consider sufficient conditions.

\begin{figure}
 \centering
 \includegraphics[width=8.5cm]{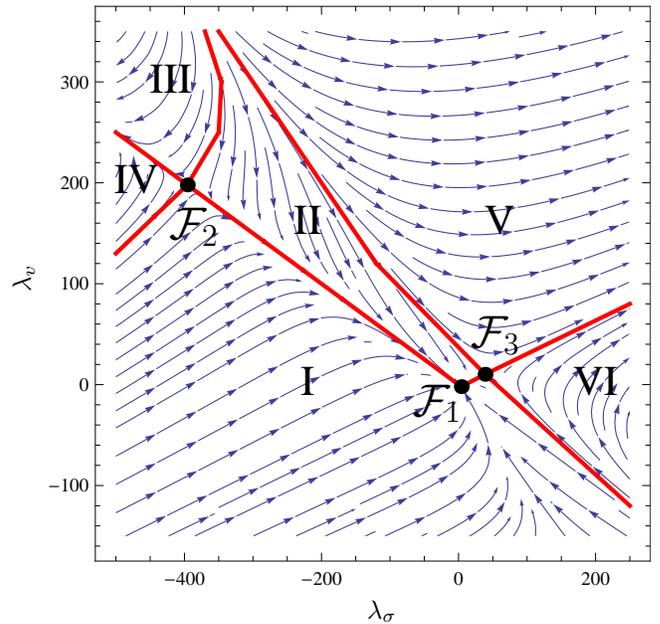}
 \caption{Phase diagram for the couplings $(\lambda_\sigma,\lambda_v)$
   for the values $(g,\lambda_3,\mu)=(1,0.25,-0.5)$ and vanishing
   anomalous dimensions $\eta_\phi=0$, using a Litim-type shape
   function, \cite{Litim:2000ci}. The arrows point towards the
   \IR. The black dots denote the fixed points, $\mathcal{F}_1$,
   $\mathcal{F}_2$ and $\mathcal{F}_3$ of the flow. All trajectories
   starting in the regions $I$ and $II$ flow towards the fully \IR{}-attractive
   (almost) Gaussian fixed point $\mathcal{F}_1$.}
   \label{f:stream}
\end{figure}

We now investigate the correspondence between the annihilation of
fixed points and well-defined \UV{}- and \IR{}-limits of the theory
for the full system $(\beta_{\lambda_\sigma},\beta_{\lambda_v})$. In
\autoref{f:stream} the phase diagram of the latter is shown for some
hypothetical \UV{}-values of the theory with
$(g,\lambda_3,\mu)\neq(0,0,0)$.
The arrows represent the flow towards the \IR. The values for $g$,
$\lambda_3$ and $\mu$ are set to $(g,\lambda_3,\mu)=(1,0.25,-0.5)$ to
demonstrate the generic behavior of the flow of
$(\lambda_\sigma,\lambda_v)$ for non-trivial values of the
gravitational couplings. The fermion anomalous dimensions are set to
zero for the plot, $\eta_\phi=0$. However, the general case
$\eta_\phi\neq0$ is considered in the rest of the work. For the
computation of the phase diagram, we use a Litim-type shape function
$\sqrt{x}\,r^{\psi}_k(x)=(1-\sqrt{x})\theta(1-x)$. The \NJL{}-model
admits three fixed points $\mathcal{F}_1$, $\mathcal{F}_2$ and
$\mathcal{F}_3$. For $g=0$, $\mathcal{F}_1$ is the Gaussian fixed
point. However, in direct analogy to the picture given for the reduced
system in \autoref{f:lam_parab}, the gravitational interaction shifts
$\mathcal{F}_1$ slightly to non-zero values. The fixed point
$\mathcal{F}_1$ is the only one which is fully \IR{}-attractive. The
other ones, namely, $\mathcal{F}_2$ and $\mathcal{F}_3$ both exhibit
one attractive and one repulsive direction. The fixed point structure
of the \NJL{}-model divides the phase diagram in \autoref{f:stream}
into six different regimes that are separated by separatrices marked
in red. Trajectories starting in the regimes $\text{I}$ and
$\text{II}$ end up in $\mathcal{F}_1$. The fact that the model
exhibits only three instead of four fixed points is a curiosity of the
\NJL{}-model in $d=4$ spacetime dimensions \cite{Braun:2011pp}. One
can think of the fourth fixed point as being located at infinity at
the point where the separatrices that separate the regimes III and V
from II, respectively, intersect.

In contrast to the simplified model of one flow equation discussed
above, for the complete \NJL{}-model the non-vanishing of fixed points
is \textit{not} a sufficient criterion for the existence of well
defined \UV{}- and \IR{}-limits any more. Due to the increased
dimensionality of the phase space $(\lambda_\sigma,\lambda_v)$ the
fixed points can lose their \UV{}- and \IR{}-attractivities,
respectively \textit{without} simultaneous fixed-point annihilation.
Hence, it is feasible that the fixed points persist but their \UV{}
and \IR{} attractiveness is flipped by the gravitational
interaction. If, say, the formerly Gaussian fixed point
$\mathcal{F}_1$ loses its \IR{}-attractivity in at least one
direction, this changes the topology of the resulting phase diagram,
possibly spoiling the well-defined \IR{}-limit of the theory.

Hence, the analysis of the phase diagram boils down to two
steps. First, we analyse, whether interacting gravity leads to an
annihilation of fixed points. Leaving the gravity-dependent quantities
$(\mathfrak{f},\mathfrak{g},\eta_\psi)$ as external input parameters
we will show, that the annihilation of fixed points is impossible for any
combination of $(\mathfrak{f},\mathfrak{g},\eta_\psi)$, once the constraints
\eqref{e:fsigns} are imposed.

In the subsequent second step, we analyse, how the fixed points's
eigenvalues, i.e., their attractivity, change under the influence of
gravitational interactions. In particular, we are interested if the
signs of the latter can change for certain combinations of
$(\mathfrak{f},\mathfrak{g},\eta_\psi)$. We will find that the signs
of the eigenvalues crucially depend on the sign of the sum
$\mathfrak{F}=1+\eta_\psi(0)+\mathfrak{f}$. This sign, however, is not
fixed by \eqref{e:fsigns}. Therefore, we perform a detailed analysis
of the flow and investigate in which regime of the relevant phase
space, $\mathfrak{F}$ does change sign. For this analysis we reduce
the system by identifying $\lambda_3=-\frac12 \mu$.  Imposing certain
regulator constraints on the anomalous dimensions $\eta_\psi$ and
$\eta_h$ we will be able to show that a change of sign of
$\mathfrak{F}$ lies either outside the physical regime of the reduced
phase space $(g,\mu)$ or outside the regime where the truncation is
reliable.

From this two-step analysis we will draw the conclusion that
gravity-induced chiral symmetry breaking is absent in interacting
4-fermion models of the \NJL{}-type.

\subsection{Fixed-Point Annihilation}
In order to study the fixed
points for the \NJL{}-model in full generality, we solve the fixed
point equation, eq.\,$\eqref{e:lamsys} = 0$, for
$(\lambda_\sigma,\lambda_v)$. Remarkably, the fixed points can be
given in a simple, closed form which reads
\begin{align}\label{e:NJLfp}
  \mathcal{F}_{1,2} &=(\frac{2\mathfrak{F}\mp\sqrt{2
      \mathfrak{g}\mathfrak{h}+4 \mathfrak{F}^2}}{\mathfrak{h}},
  \frac{-2\mathfrak{F}\pm\sqrt{2 \mathfrak{g}\mathfrak{h}+
      4 \mathfrak{F}^2}}{2 \mathfrak{h}})	\,,	\notag\\
  \mathcal{F}_3 &=(-\frac{2
    \mathfrak{g}\mathfrak{h}+3\mathfrak{F}^2}{8
    \mathfrak{F}\mathfrak{h}},-\frac{-2
    \mathfrak{g}\mathfrak{h}+\mathfrak{F}^2}{8
    \mathfrak{F}\mathfrak{h}})\,,
\end{align}
where, again, $\mathfrak{F}=(1+\eta_\psi(0)+\mathfrak{f})$. Note, that
the fixed points of the \NJL{}-model without gravity are recovered by
setting $g\to0$ and, correspondingly, $\mathfrak{F}\to1$ and
$\mathfrak{g}\to0$. In this case, $\mathcal{F}_1\to(0,0)$ which means
that it is the Gaussian fixed point in this limit. Clearly, the fixed
points $\mathcal{F}_1$ and $\mathcal{F}_2$ annihilate, if the argument
in the square roots of the first line in \eqref{e:NJLfp} becomes
negative. The third fixed point $\mathcal{F}_3$ is present for any
combination $(\mathfrak{f},\mathfrak{g},\eta_\psi)$. Thereby, the
condition for the annihilation of fixed points translates into the
simple inequality
\begin{align}\label{e:cond}
  \underbrace{\mathfrak{g}\mathfrak{h}}_{>0}+\underbrace{2
    (1+\eta_\psi(0)+\mathfrak{f})^2}_{\geq0}\overset{!}{<}0\,.
\end{align}
Due to the fixed signs of
$\mathfrak{g}$ and $\mathfrak{h}$ given by the equations
\eqref{e:fsigns}, the product $\mathfrak{gh}$ is always positive. This
suggests that neither of the two terms in \eqref{e:cond} becomes
negative. Therefore, the inequality \eqref{e:cond} is not fulfilled
for any choice of regulator.

In particular, the fermion anomalous dimension at vanishing momentum
$\eta_\psi(0)$ as well as $\mathfrak{f}$ do not play any r\^{o}le for the
sign of left-hand side of \eqref{e:cond} since they enter
quadratically. We have shown that the annihilation of fixed points due
to gravitational interactions in our model is impossible. The only
possible way to destroy the well-defined limits now, is via the change
of stability of the fixed points.

\subsection{Stability of Fixed Points}
We now turn to our second
criterion, the stability of the fixed points in the \NJL{}-model. It
is conceivable, that, if the gravitational interaction is strong
enough, it might change the attractivity of its fixed points. In the
vicinity of fixed points, the flow can be linearised. Thus, its
behaviour is described by critical exponents $\theta_n$ which are the
eigenvalues of the matrix $\left(\frac{\partial
    \beta_{\lambda_i}}{\partial \lambda_j}\right)$ to wit
\begin{align}
  \theta_n=\mbox{Spec}\left[\left(\frac{\partial
        \beta_{\lambda_i}}{\partial
        \lambda_j}\right)\bigg|_{\mathcal{F}_n}\right]\,.
\end{align}
Hence, from the flow equations \eqref{e:lamsys}, together with the
fixed point values \eqref{e:NJLfp} we find the critical exponents of
the \NJL{}-model with gravity. They take the simple form
\begin{align}\label{e:thetas}
\begin{split}
  \theta_{1}&=\left\{6 \mathfrak{F}-4 \sqrt{\mathfrak{F}^2+
      \frac{\mathfrak{h}\mathfrak{g}}{2}},2\sqrt{
      \mathfrak{F}^2+\frac{\mathfrak{h}\mathfrak{g}}{2}}\right\}\,,	\\
  \theta_2&=\left\{6 \mathfrak{F}+4 \sqrt{
      \mathfrak{F}^2+\frac{\mathfrak{h}\mathfrak{g}}{2}},
    -2 \sqrt{\mathfrak{F}^2+\frac{\mathfrak{h}\mathfrak{g}}{2}}\right\}\,,	\\
  \theta_3&=\left\{\frac{\mathfrak{F}^2+2
      \mathfrak{h}\mathfrak{g}\pm\sqrt{81 \mathfrak{F}^4+4
        \mathfrak{h}\mathfrak{g}(\mathfrak{h}
        \mathfrak{g}-7\mathfrak{F}^2)}}{4 \mathfrak{F}}\right\}\,.
\end{split}
\end{align}
Considering again the limit of vanishing gravitational coupling
$g\to0$, thus $\mathfrak{F}\to1$ and $\mathfrak{g}\to0$, we get
$\theta_1=\{2,2\}$, $\theta_2=\{10,-2\}$ and $\theta_3=\{5/2,-2\}$
which we know from the \NJL{}-model without gravity. Note that these
quantities are regulator independent in the limit $g\to0$, since the
fermion anomalous dimension does not depend on the 4-fermion
couplings. Hence, this fact is a consequence of the momentum
independent 4-fermion interactions considered here. For $g\neq 0$
however, the regulator dependent quantities $\mathfrak{F}$,
$\mathfrak{h}$ and $\mathfrak{g}$ enter the equations
\eqref{e:thetas}. From these equations we can see that na\"{i}vely,
the $\theta_i$ could flip sign, if $\mathfrak{F}$ becomes
negative. Since $\mathfrak{F}=1+\eta_\psi(0)+\mathfrak{f}$ and neither
$\eta_\psi(p^2)$ nor $\mathfrak{f}$ have a definite sign, the case
$\mathfrak{F}<0$ is not excluded generally. However, since
$(\eta_\psi,\mathfrak{f})\to(0,0)$ for $g\to0$, the gravitational coupling
$g$ has to be strong in order to allow for $\mathfrak{F}<0$. Hence,
due to the generic flow of $g$ depicted in \autoref{fig:al_k} (right panel) this is
the case only in the deep \UV.
\subsection{The sign of $\mathfrak{F}$}
The sign of $\mathfrak{F}$ in
the deep \UV{} is not fixed for general regulators. However, we will show
that definite statements about the latter are indeed
possible once a regulator is chosen. In the following, we consider
$\sqrt{x}\,r^{\psi}_k(x)=(1-\sqrt{x})\theta(1-x)$ and
$x\,r_k^h(x)=(1-x)\theta(1-x)$ which allow for analytic flow
equations. For these choices, $\mathfrak{F}$ is given by
\begin{align}\label{eq:F}
 \mathfrak{F}=\frac{g}{70\pi} \left(\frac{448-79 \eta_h(k^2)}{8 (1
+\mu)^2}-\frac{7-2 \eta_\psi(k^2)}{1+\mu}\right)+\eta_\psi(0)+1\,,
\end{align}
where we have moved the anomalous dimensions out of the loop integrals
at momenta at which the integrand is peaked, i.e. $q^2=k^2$. This is
shown to be a good approximation in
\cite{Christiansen:2014raa,Meibohm:2015twa}.  Now we investigate, for
which values of $1+\mu$, $\mathfrak{F}$ does become zero, and, thus
where $\mathfrak{F}<0$ is in principle possible. For the following
analysis, we make use of the bounds of the anomalous dimensions
$\eta_\psi(p^2)<1$ and $\eta_h(p^2)<2$, which are necessary such that
the corresponding class of regulators, \eqref{e:regclass}, remains
well-behaved, for details see \cite{Meibohm:2015twa}. We will take
$\eta_\psi(p^2)$ as momentum-independent and write
$\eta_\phi=\eta_\phi(p^2)$. Furthermore, we assume
that $g$ does not take large values $g\lessapprox10$ along the flow
from $k=\infty$ to $k\to0$. This assumption is supported by the \FRG{}
studies of pure quantum gravity and quantum gravity with the given
matter content $N_f=1$ known to the authors. Summarising the main
steps, we solve $\mathfrak{F}=0$ for $1+\mu$, where $\mathfrak{F}$
given in \eqref{eq:F}. Employing the approximation for
$\eta_\psi(p^2)=\eta_\psi$ and the regulator constraints first leads
to
\begin{align}\label{eq:etapsilim}
	-1<\eta_\psi<1\,.
\end{align}
The assumption that $g$ does not become large, implies that
$\eta_\psi$ must be very close to the lower limit in
\eqref{eq:etapsilim}.  We express this finding as
 \begin{align}\label{eq:delta}
  1+\eta_\psi=:\delta\ll1\,,
 \end{align}
 which allows to expand the equation for $1+\mu$ in a Taylor series in
 $\delta$. More details can be found in Appendix \ref{a:fsign}. The
 analysis results in a lower bound for $1+\mu$, namely
\begin{align}\label{eq:mubound}
 1+\mu>\frac{145}{36}>4\,.
\end{align}
Since all analyses of the \UV{} behavior of quantum gravity with
non-zero $\eta_\psi$ report \UV-values of $1+\mu$ considerably smaller
than 4, we conclude that the condition \eqref{eq:mubound} is not
fulfilled in the \UV. In the \IR, where \eqref{eq:mubound} can hold,
\eqref{eq:delta} is not fulfilled, since $\eta_\psi \ll 1$ and,
thus, $\delta\approx1$.

Truncations with zero fermion anomalous dimension $\eta_\psi=0$
\cite{Percacci:2002ie,Eichhorn:2011pc} require very large values of
$g$ of the order of $10^2$ in order to flip the sign of
$\mathfrak{F}$, which we regard as unphysical.

In summary, we find that the
flipping of signs for the critical exponents $\theta_i$ does not take
place, as long as certain consistency conditions for the anomalous
dimensions and the size of $g$ are met. Thus, the flipping of signs of
the critical exponents lies within a region in parameter space which
is not reached by the flow. We conjecture, that this is also true for
more general classes of regulators. This, however, can presumably be
verified only numerically.
\section{Multiple Fermions}
In this section we analyse whether the mechanism that prevents chiral
symmetry breaking in the 4-fermion system is an artifact of the
fermion number $N_f=1$ discussed above. In order to understand the
impact of $N_f$ on the previous results, we consider a Fierz-complete
model with $N_f$ fermions and $\text{SU}(N_f)\times\text{SU}(N_f)$
chiral symmetry. This model was also studied in a similar spirit in
\cite{Eichhorn:2011pc}. There, chiral symmetry breaking was tested
with a fermion anomalous dimension that was treated non-dynamically,
as an input parameter. Here, we put forward a complete analysis of the
fixed point annihilation with dynamical anomalous dimensions and
general regulators.  The Fierz-complete fermion potential $V_{N_f}$ is
given by
\begin{multline}\label{e:nfac}
  V_{N_f}(\psi,\bar \psi) = -\bar\lambda_{-}\left[(
    \bar\psi^i\gamma_\mu\psi^i)^2+(\bar\psi^i
    \gamma_\mu\gamma_5\psi^i)^2\right]\\
  -\bar\lambda_{+}\left[(\bar\psi^i\gamma_\mu\psi^i)^2
    +(\bar\psi^i\gamma_\mu\gamma_5\psi^i)^2\right]\,.
\end{multline}
This model exhibits four fixed points. For $g=0$ one of them is in the
fully \IR{}-attractive Gaussian fixed point. The second and third
fixed points exhibit one \IR{}-attractive and one repulsive
direction. The fourth fixed point is fully \IR{}-repulsive. Along the
same lines as before we derive conditions for chiral symmetry breaking
from a two step procedure: We first analyse the possibility of fixed
point annihilation followed by a brief analysis
of their stability.
\subsection{Fixed Point Annihilation}
The flow equations for the present
$\text{SU}(N_f)\times\text{SU}(N_f)$ chirally symmetric \NJL{}-type
interacting fermion model, \eq{eq:flownf}, are very similar to the
previous ones.  The purely fermionic part agrees with
\cite{Gies:2003dp,Eichhorn:2011pc} and, when employing the Litim-type
regulators given above, with \cite{Eichhorn:2011pc}. However, due to a
slightly different choice of the graviton-gauge ($\beta=1$ here, and
$\beta=0$ in \cite{Eichhorn:2011pc}), the contributions involving
gravity interactions differ slightly. The $\beta$-functions are given
by are given by
\begin{widetext}
 \begin{align}\label{eq:flownf}
  \begin{split}
    \beta_{\lambda_-} &= -2(N_f-1)\mathfrak{h}\lambda
    _-^2+2(1+\eta_\psi(0)+ \mathfrak{f})\,\lambda_-
    -2N_f\,\mathfrak{h}\lambda_+^2
    +\frac14\mathfrak{g}\,,	\\
    \beta_{\lambda_+} &= -6\mathfrak{h}\lambda
    _+^2+2(1+\eta_\psi(0)+\mathfrak{f}-2(N_f+1)\,
    \mathfrak{h}\lambda_-)\lambda_+ -\frac14\mathfrak{g}\,.
 \end{split}
 \end{align}
 \end{widetext}
We observe that in the model with $N_f$ fermions, the fully \IR{}-attractive
Gaussian fixed point potentially annihilates with one of
 the semi-stable fixed points, which was also true for the
 \NJL{}-model discussed above. Additionally, the new fully
 \IR{}-repulsive (fully \UV{}-attractive) fixed point potentially
 annihilates with the other semi-stable fixed point. In both cases,
 the annihilation of the respective fixed points changes the topology
 of the phase diagram and possibly spoils the well-defined \UV{}- and
 \IR{}-limits. This potentially leads to chiral symmetry breaking
 during the flow. Remarkably, the fixed point equation,
 eq.\,$\eqref{eq:flownf}=0$, can, again, be solved in a simple closed
 form which is given in Appendix \ref{a:fpnf}.  From the arguments of
 the corresponding square roots we derive two inequalities that
 are characteristic for the annihilation of the fixed points
 involved. They are given by

 \begin{align}\label{e:cond2}
 \begin{split}
   \underbrace{\mathfrak{gh}(2 N_f-1)}_{>0}+\underbrace{2
     \mathfrak{F}^2}_{\geq0}	&\overset{!}{<}	0\,,\\
   \underbrace{\mathfrak{gh}\mathfrak{G} (N_f-1)}_{\geq0}+2
   \underbrace{\mathfrak{F}^2 (N_f+3)^2}_{\geq0} &\overset{!}{<} 0\,,
 \end{split}
 \end{align}
where $\mathfrak{G} = (9 + 4N_f + 3N_f^2)>0$. With the same arguments
as before it can be verified, that the two inequalities
\eqref{e:cond2}, can both not be fulfilled individually for any choice
of $N_f\geq1$. Thus, also in this case the annihilation of fixed
points is avoided on general grounds by the structure of the
4-fermion interaction.
\subsection{Stability of Fixed Points}
The stability of the fixed points in the general case of $N_f$
fermions turns out to be very extensive and goes beyond the scope of
this work. Therefore we will only consider the limiting cases
$N_f\approx1$ and $N_f\ra\infty$. The former case is identical to the
analysis of the \NJL{}-model with one fermion discussed above, where
we noted the significance of the sign of $\mathfrak{F}$. Hence, we
now study the stability of the fixed point of the system with $N_f$
fermions in a large-$N_f$-expansion. In \cite{Meibohm:2015twa} the
minimally-coupled gravity-fermion system has been studied.  Let us
first consider the approximation where the contributions of the
anomalous dimensions are neglected. Then, the minimally-coupled sub-system
reveals an asymptotic scaling of the \UV{} fixed point value of $g$,
$g^*$, as $g^*\sim1/N_f$ for $N_f\to\infty$, \cite{Meibohm:2015twa}.
This motivates a rescaling of the couplings according to
\begin{align}
  (g,\lambda_i) \rightarrow (\tilde g= N_f g, \tilde \lambda_i = N_f
  \lambda_i)\,,
\end{align}
such that $\tilde g^*$ remains constant as $N_f\to\infty$.
Expanding the fixed points for the resulting
system in $1/N_f$ we find 
{\small 
\begin{align}\nonumber 
  \mathcal{\tilde F}_{1/2} \to \bigg\{&\hspace{-2pt}\mp\hspace{-2pt}\frac{\mathfrak{\tilde
      g}}{8 N_f}\,,\,\frac{1}{2\mathfrak{h}}
  \left(1+\frac{1\hspace{-2pt}+\hspace{-2pt}\mathfrak{f}}{N_f}
    \mp\left(1+\frac{4(1+\mathfrak{f})\hspace{-2pt}+\hspace{-2pt}\mathfrak{h\tilde
          g}}{4N_f}\right)\right)
  \bigg\}	\,,\\[2ex]
  \mathcal{\tilde F}_{3/4} \to \bigg\{&
 \hspace{-2pt}\mp\hspace{-2pt}\frac{1}{2\mathfrak{h}}\left(1+\frac{2\hspace{-2pt}+\hspace{-2pt}4 \mathfrak{f}\hspace{-2pt}+\hspace{-2pt}
      \mathfrak{h \tilde g}}{4 N_f}\right)\,,\,
  \frac{1}{2\mathfrak{h}}\left(1-\frac{1\hspace{-2pt}+\hspace{-2pt}\mathfrak{f}}{N_f}\pm\frac{6\hspace{-2pt}+\hspace{-2pt}\mathfrak{h
        \tilde g}}{4 N_f}\right)\bigg\} \,, \label{e:nfscal}
\end{align}
}
where $\mathfrak{\tilde g}=\mathfrak{g}(\tilde
g)=N_f^2\,\mathfrak{g}(g)$, see eq.\,\eqref{e:fsigns}. For
$N_f\to\infty$ the gravity contributions in \eq{e:nfscal} tend towards
zero. Then, only diagrams with purely fermionic loops contribute to
the flow. All gravity contributions are sub-leading and the flow of
the 4-fermion couplings is independent of the rest of the flow. In
this limit we arrive at the fixed points $\mathcal{\tilde
  F}_{1/2}\sim\{0,\frac1{2\mathfrak{h}}(1\mp1)\}$ and $\mathcal{\tilde
  F}_{3/4}\sim\{\mp\frac1{2\mathfrak{h}},\frac1{2\mathfrak{h}}\}$. Note that
  $\mathcal{\tilde F}_1$ is the Gaussian fixed point here. The critical exponents also reflect this gravitational
decoupling, and are given in Appendix \ref{a:largenf} to the order
$\mathcal{O}(1/N_f)$. Hence, the scaling $g^*\sim1/N_f$ leads to the
expected behaviour in the large
$N_f$-limit: the large number of fermions dominates the fluctuation
physics.

However, the situation is more complicated in the full system with
momentum-dependent anomalous dimensions. Then, the minimally-coupled
sub-system exhibits an asymptotic scaling of approximately
$g^*\sim1/\sqrt{N_f}$, \cite{Meibohm:2015twa}, in contradistinction to
the $1/N_f$ discussed above. In this case, the large-$N_f$-limit is
non-trivial and fingerprints of the gravitational interactions survive
the fermionic dominance. This is readily verified by replacing
$\mathfrak{\tilde g} \to N_f \mathfrak{\tilde g}$ in \eqref{e:nfscal}
in order to account for the modified asymptotic $N_f$-scaling.  This
leads to 
\begin{align}\nonumber 
\mathcal{\tilde F}_{1/2}&\to \left\{\mp\frac{\mathfrak{\tilde
    g}}{8}\,,\,\frac1{2\mathfrak{h}}\left(1\mp\left(1+\frac{\mathfrak{h \tilde
    g}}{4}\right)\right)\right\}\,,\\[2ex] 
\mathcal{\tilde
  F}_{3/4}& \to  \left\{\mp\frac1{2\mathfrak{h}}\left(1+\frac{\mathfrak{h \tilde
    g}}{4}\right)\,,\,\frac1{2\mathfrak{h}}\left(1\pm\frac{\mathfrak{h \tilde
    g}}{4}\right)\right\}\,, 
\end{align}
in the limit $N_f\to \infty$.  A more detailed analysis of the
$g^*\sim1/\sqrt{N_f}$-scaling reveals that these remainders of gravity
do, however, not allow for chiral symmetry breaking in the
large-$N_f$-limit.

This discussion shows that the scaling of $g^*$ for a large number of
matter fields is of crucial importance for the properties of the
combined theory. In particular, it was found in \cite{Meibohm:2015twa}
that the fixed point value $g^*$ grows as the number of scalar fields,
$N_s$, increases, which is in sharp contrast to the fermionic case
discussed here. It is therefore tempting to study the flow of scalar
self-interactions (see also \cite{Eichhorn:2012va}), at moderate $N_s$
taking into account the corresponding scaling of $g^*$.

The current analysis suggests, that gravity plays no or only a
negligible r\^{o}le in the large-$N_f$-limit and chiral symmetry
breaking is absent. Coming back to our original problem, we conclude
that the sign of $\mathfrak{F}$ is of interest only at low and
intermediate $N_f$, where the gravitational contributions are not
suppressed by fermion fluctuations. Only in this regime can negative
$\mathfrak{F}$ spoil the stability of the fixed points. Such a
negative $\mathfrak{F}$ requires $(1+\mu)>4$, as discussed
above. However, in \cite{Meibohm:2015twa} it has been shown that
$1+\mu$ stays well below this limit for all $N_f$ and approaches 0 as $N_f\to\infty$. We
conclude that chiral symmetry breaking is absent not only in the present
combination of a general interaction 4-fermion model and the asymptotically-safe
minimally-coupled gravity-fermion sub-system put forward in \cite{Meibohm:2015twa}.
Instead this analysis implies that all combinations of the present 4-fermion model
together with arbitrary minimally-coupled gravity-fermion sub-systems avoid chiral symmetry
breaking, provided these sub-systems exhibit the following properties: i) They are asymptotically
safe. ii) They assure $1+\mu<4$ along the whole \RG-trajectory. iii) They allow for a large-$N_f$-limit where
$g^*\to0$ as $N_f\to\infty$ with a scaling $g^*\sim1/\sqrt{N_f}$ or faster.

\section{Conclusions}
We have analysed the possibility of chiral symmetry breaking in
theories of interacting fermions and quantum gravity using a
self-consistent \FRG-approach. As a general feature for models with
4-fermion interaction of the \NJL-type we found that chiral symmetry
breaking is absent irrespective of the number of fermion fields and
mostly independent of the regularisation scheme.

For this analysis, we have reduced the condition for the existence of
well defined \UV{}- and \IR{}-limits to the existence and stability of
fixed points of the renormalisation group flow. We have interpreted
the possible annihilation as well as the change of stability of the
latter as indications for a topological change in the phase diagram of
the 4-fermion couplings, which could lead to chiral symmetry breaking
in the \UV. We have found that fixed point annihilation is ruled out
generally for the models discussed here independent of the chosen
regulator and the fermion number. However, the change of attractivity
of the fixed points is more subtle in particular for an arbitrary
number of fermions.  Still, we were able to argue that in neither of
the present models we expect the fixed-point-attractivity to change
due to the interacting of interacting fermions with gravity. This
conclusion is drawn from a more detailed analysis based on a specific
choice of regulator in the limits $N_f\approx1$ and $N_f\to\infty$. In
the latter case, the flow of the 4-fermion couplings decouples either
completely or at least mostly from the rest of the fermion-gravity
system. The degree of the latter decoupling depends on the scaling of
the fixed point value $g^*$ as $N_f\to\infty$. Our results imply that
gravity-induced chiral symmetry breaking at the Planck scale is
avoided for a general class of models with chirally symmetric
4-fermion interactions.

\acknowledgements

We thank N.~Christiansen, A.~Eichhorn, H.~Gies, M.~Reichert,
M.~Scherer and C.~Wetterich for discussions. This work is supported by
EMMI and by ERC-AdG-290623.

\appendix
\section{sign of $\mathfrak{F}$ as function of $\mu$}\label{a:fsign}
In the following, we analyse in detail the sign of
$\mathfrak{F}=1+\eta_\psi(0)+\mathfrak{f}$ for a Litim-type
regulator. The explicit expression is given in \eqref{eq:F}.  For
vanishing gravitational constant $g\to0$ we have
$\mathfrak{F}\to1$. Hence for small values of $g$ we will have
$\mathfrak{F}\approx 1$.  We see, that some of the critical exponents
given in \eqref{e:thetas} change sign as soon as $\mathfrak{F}$
becomes negative. We now solve $\mathfrak{F}=0$ for $1+\mu$ for
momentum independent fermion anomalous dimensions $\eta_\psi(p^2)= \eta_\psi$.
For the given class of regulators,
\eqref{e:regclass}, the $\eta$'s have to satisfy the bounds
$\eta_\psi<1$ and $\eta_h<2$.  Solving eq.\,$\eqref{eq:F}=0$ we get
\begin{align}\nonumber 
&  (1+\mu)_{1,2} = \frac1{140 \pi  (1+\eta_\psi)}\bigg((7-2 \eta_\psi) g \\
&  \left.\pm\sqrt{g \left(35 \pi (1+\eta_\psi) (79 \eta_h-448)+(7-2
        \eta_\psi)^2 g\right)}\right)\,.
\label{eq:1pmu}\end{align}
This equation only has real solutions if the argument in the square
root is positive-semidefinite. Hence, in order to approach the limit
$\mathfrak{F}\to0$ for a real $\mu$ we need
\begin{align}\label{eq:sqrtarg}
  (7-2 \eta_\psi)^2 g\geq35 \pi (1+\eta_\psi)
  (448-79 \eta_h)\,.
\end{align}
The left-hand side is positive-semidefinite. Hence, we see that for
$\eta_\psi\leq-1$ the above inequality is satisfied for
$\eta_h<2$. However, $1+\eta_\psi\to0$ leads to a solution of
\eqref{eq:1pmu} where $1+\mu\to\infty$. This means that the flow would need
to pass the graviton-propagator pole in order to reach the regime
$\eta_\psi<-1$. This transition, however, is not admitted by the flow equations.
Therefore, we conclude $-1<\eta_\psi<1$. Then, for
$\eta_h<2$ we find that the right-hand side of \eqref{eq:sqrtarg} is
positive, and we write
\begin{align}\label{eq:gcond}
  g>35 \pi (448-79 \eta_h) \frac{1+\eta_\psi}{ (7-2
    \eta_\psi)^2}>10150 \pi \frac{1+\eta_\psi}{ (7-2 \eta_\psi)^2}\,.
\end{align}
Due to $-1<\eta_\psi<1$ we have
\begin{align}
	0<\frac{1+\eta_\psi}{ (7-2 \eta_\psi)^2}<\frac2{25}\,.
\end{align}
Typical values of $g$ in the \UV{} are of order 1, and we restrict
ourselves to couplings $g\lessapprox10$. The numerator $(1+\eta_\psi)$
in \eqref{eq:gcond} has to become very small and, in particular,
considerably smaller than $g$. This is expressed as
\begin{align}
 (1+\eta_\psi)=:\delta\ll1\,.
 \end{align}
 For small $\delta$ we can expand \eqref{eq:1pmu} in $\delta$ and get,
 keeping only the lowest orders in $\delta$,
\begin{align}
  (1+\mu)_1	&\approx	\frac{9 g}{70 \pi  \delta }\,,	\\[2ex]
  (1+\mu)_2 &\approx \frac{56}{9}-\frac{79
    \eta_h}{72}>\frac{145}{36}>4\,.
\end{align}
We have used $\eta_h<2$ to arrive at the lower bound for $(1+\mu)_2$.
Now, using equation \eqref{eq:gcond} and expanding in $\delta$ we get
a lower bound for $(1+\mu)_{1}$, which reads
\begin{align}
  (1+\mu)_{1}>140\,,
\end{align}
to lowest order in $\delta$. Thus, respecting the regulator bound, we
must have
\begin{align}
 (1+\mu)>4\,,
\end{align}
in order for $\mathfrak{F}$ to become negative.
\\

\section{Fixed Points for $N_f$ Fermions}\label{a:fpnf}
The fixed-point equation, eq.\,$\eqref{eq:flownf}=0$, can be solved in
a simple and closed form for $(\lambda^*_-,\lambda^*_+)$ for an
arbitrary number of fermions. The solutions are given by the four
fixed points $\mathcal{F}^{(N)}_{1\leq i\leq4}$ to wit
{\small
\begin{widetext}
	\begin{align}
	\begin{split}
          \mathcal{F}^{(N)}_{1/2}=&\left(-\frac{2
              \mathfrak{F}\mp\sqrt{4 \mathfrak{F}^2+2 (2N_f-1)
                \mathfrak{h}\mathfrak{g}}}{4 (2N_f-1)
              \mathfrak{h}},\frac{\mp2 \mathfrak{F}+\sqrt{4
                \mathfrak{F}^2+2 (2N_f-1) \mathfrak{h}\mathfrak{g}}}{4
              (2N_f-1) \mathfrak{h}}\right)\,,\\[2ex] 
          \mathcal{F}^{(N)}_{3/4}=&\left(\frac{2
              \mathfrak{F}(N_f+3)\pm\sqrt{4 (N_f+3)^2 \mathfrak{F}^2+
                2 (N_f-1) \mathfrak{h}\mathfrak{g} \mathfrak{G}}}{4
              \mathfrak{h}\mathfrak{G}}, \right.  
          \left. \frac{2
              \mathfrak{F}(\mathfrak{G}+(N_f-1) N_f)\mp(N_f+3) \sqrt{4
                (N_f+3)^2 \mathfrak{F}^2+ 2 (N_f-1)
                \mathfrak{h}\mathfrak{g} \mathfrak{G}}}{4
              \mathfrak{h}\mathfrak{G}(N_f-1)}\right)\,.
        \end{split}
	\end{align}
\end{widetext}
}
Without gravity (where $g=0$), we have $(\mathfrak{f},\mathfrak{g})\to(0,0)$
and hence $\mathcal{F}_1^{(N)}\ra(0,0)$ is the Gaussian fixed
point. For $g\neq0$ all fixed points are non-Gaussian.
\section{Critical Exponents in the Large-$N_f$-Limit}\label{a:largenf}
The critical exponents for the rescaled couplings $(\tilde
\lambda_\sigma,\tilde \lambda_v)$ for large $N_f$ take the form
\begin{widetext}
\begin{align}
	\begin{split}
  \tilde \theta_{1/3} &\approx \left\{\pm(2+\frac{2
      \mathfrak{\tilde f}}{N_f}),2+\frac{2+2 \mathfrak{\tilde f}+ \mathfrak{h}
      \mathfrak{\tilde g}}{N_f}\mp\frac{2}{N_f}\right\} \,, \qquad \tilde
  \theta_{4} \approx \left\{-2 -\frac{2 \mathfrak{\tilde f}+\mathfrak{h}
      \mathfrak{\tilde g}}{N_f},2
    +\frac{2 \mathfrak{\tilde f}+4}{N_f}\right\}\,,	\\[2ex] 
  \tilde \theta_{2} &\approx \left\{-2-\frac{8+4
      \mathfrak{\tilde f}+\mathfrak{h} \mathfrak{\tilde g}+\sqrt{64-\mathfrak{h}^2
        \mathfrak{\tilde g}^2}}{2 N_f},-2-\frac{8+4 \mathfrak{\tilde f}+\mathfrak{h}
      \mathfrak{\tilde g}-\sqrt{64-\mathfrak{h}^2 \mathfrak{\tilde g}^2}}{2
      N_f}\right\}\,,
	\end{split} 
\end{align}
\end{widetext}
where $\mathfrak{\tilde f}=\mathfrak{f}(\tilde g)$.  To leading order in $1/N_f$,
the gravity contributions decouple from the 4-fermion model.
\ 
\eject
\  
%

\begin{thebibliography}{51}%
\makeatletter
\providecommand \@ifxundefined [1]{%
 \@ifx{#1\undefined}
}%
\providecommand \@ifnum [1]{%
 \ifnum #1\expandafter \@firstoftwo
 \else \expandafter \@secondoftwo
 \fi
}%
\providecommand \@ifx [1]{%
 \ifx #1\expandafter \@firstoftwo
 \else \expandafter \@secondoftwo
 \fi
}%
\providecommand \natexlab [1]{#1}%
\providecommand \enquote  [1]{``#1''}%
\providecommand \bibnamefont  [1]{#1}%
\providecommand \bibfnamefont [1]{#1}%
\providecommand \citenamefont [1]{#1}%
\providecommand \href@noop [0]{\@secondoftwo}%
\providecommand \href [0]{\begingroup \@sanitize@url \@href}%
\providecommand \@href[1]{\@@startlink{#1}\@@href}%
\providecommand \@@href[1]{\endgroup#1\@@endlink}%
\providecommand \@sanitize@url [0]{\catcode `\\12\catcode `\$12\catcode
  `\&12\catcode `\#12\catcode `\^12\catcode `\_12\catcode `\%12\relax}%
\providecommand \@@startlink[1]{}%
\providecommand \@@endlink[0]{}%
\providecommand \url  [0]{\begingroup\@sanitize@url \@url }%
\providecommand \@url [1]{\endgroup\@href {#1}{\urlprefix }}%
\providecommand \urlprefix  [0]{URL }%
\providecommand \Eprint [0]{\href }%
\providecommand \doibase [0]{http://dx.doi.org/}%
\providecommand \selectlanguage [0]{\@gobble}%
\providecommand \bibinfo  [0]{\@secondoftwo}%
\providecommand \bibfield  [0]{\@secondoftwo}%
\providecommand \translation [1]{[#1]}%
\providecommand \BibitemOpen [0]{}%
\providecommand \bibitemStop [0]{}%
\providecommand \bibitemNoStop [0]{.\EOS\space}%
\providecommand \EOS [0]{\spacefactor3000\relax}%
\providecommand \BibitemShut  [1]{\csname bibitem#1\endcsname}%
\let\auto@bib@innerbib\@empty
\bibitem [{\citenamefont {Christiansen}\ \emph {et~al.}(2015)\citenamefont
  {Christiansen}, \citenamefont {Knorr}, \citenamefont {Meibohm}, \citenamefont
  {Pawlowski},\ and\ \citenamefont {Reichert}}]{Christiansen:2015rva}%
  \BibitemOpen
  \bibfield  {author} {\bibinfo {author} {\bibfnamefont {N.}~\bibnamefont
  {Christiansen}}, \bibinfo {author} {\bibfnamefont {B.}~\bibnamefont {Knorr}},
  \bibinfo {author} {\bibfnamefont {J.}~\bibnamefont {Meibohm}}, \bibinfo
  {author} {\bibfnamefont {J.~M.}\ \bibnamefont {Pawlowski}}, \ and\ \bibinfo
  {author} {\bibfnamefont {M.}~\bibnamefont {Reichert}},\ }\href {\doibase
  10.1103/PhysRevD.92.121501} {\bibfield  {journal} {\bibinfo  {journal} {Phys.
  Rev.}\ }\textbf {\bibinfo {volume} {D92}},\ \bibinfo {pages} {121501}
  (\bibinfo {year} {2015})},\ \Eprint {http://arxiv.org/abs/1506.07016}
  {arXiv:1506.07016 [hep-th]} \BibitemShut {NoStop}%
\bibitem [{\citenamefont {Meibohm}\ \emph {et~al.}(2015)\citenamefont
  {Meibohm}, \citenamefont {Pawlowski},\ and\ \citenamefont
  {Reichert}}]{Meibohm:2015twa}%
  \BibitemOpen
  \bibfield  {author} {\bibinfo {author} {\bibfnamefont {J.}~\bibnamefont
  {Meibohm}}, \bibinfo {author} {\bibfnamefont {J.~M.}\ \bibnamefont
  {Pawlowski}}, \ and\ \bibinfo {author} {\bibfnamefont {M.}~\bibnamefont
  {Reichert}},\ }\href@noop {} {\  (\bibinfo {year} {2015})},\ \Eprint
  {http://arxiv.org/abs/1510.07018} {arXiv:1510.07018 [hep-th]} \BibitemShut
  {NoStop}%
\bibitem [{\citenamefont {Weinberg}(1979)}]{Weinberg:1980gg}%
  \BibitemOpen
  \bibfield  {author} {\bibinfo {author} {\bibfnamefont {S.}~\bibnamefont
  {Weinberg}},\ }\href@noop {} {\bibfield  {journal} {\bibinfo  {journal}
  {General Relativity: An Einstein centenary survey, Eds. Hawking, S.W.,
  Israel, W; Cambridge University Press}\ ,\ \bibinfo {pages} {790}} (\bibinfo
  {year} {1979})}\BibitemShut {NoStop}%
\bibitem [{\citenamefont {Reuter}(1998)}]{Reuter:1996cp}%
  \BibitemOpen
  \bibfield  {author} {\bibinfo {author} {\bibfnamefont {M.}~\bibnamefont
  {Reuter}},\ }\href {\doibase 10.1103/PhysRevD.57.971} {\bibfield  {journal}
  {\bibinfo  {journal} {Phys. Rev.}\ }\textbf {\bibinfo {volume} {D57}},\
  \bibinfo {pages} {971} (\bibinfo {year} {1998})},\ \Eprint
  {http://arxiv.org/abs/hep-th/9605030} {arXiv:hep-th/9605030} \BibitemShut
  {NoStop}%
\bibitem [{\citenamefont {Souma}(1999)}]{Souma:1999at}%
  \BibitemOpen
  \bibfield  {author} {\bibinfo {author} {\bibfnamefont {W.}~\bibnamefont
  {Souma}},\ }\href {\doibase 10.1143/PTP.102.181} {\bibfield  {journal}
  {\bibinfo  {journal} {Prog.Theor.Phys.}\ }\textbf {\bibinfo {volume} {102}},\
  \bibinfo {pages} {181} (\bibinfo {year} {1999})},\ \Eprint
  {http://arxiv.org/abs/hep-th/9907027} {arXiv:hep-th/9907027 [hep-th]}
  \BibitemShut {NoStop}%
\bibitem [{\citenamefont {Reuter}\ and\ \citenamefont
  {Saueressig}(2002)}]{Reuter:2001ag}%
  \BibitemOpen
  \bibfield  {author} {\bibinfo {author} {\bibfnamefont {M.}~\bibnamefont
  {Reuter}}\ and\ \bibinfo {author} {\bibfnamefont {F.}~\bibnamefont
  {Saueressig}},\ }\href {\doibase 10.1103/PhysRevD.65.065016} {\bibfield
  {journal} {\bibinfo  {journal} {Phys. Rev.}\ }\textbf {\bibinfo {volume}
  {D65}},\ \bibinfo {pages} {065016} (\bibinfo {year} {2002})},\ \Eprint
  {http://arxiv.org/abs/hep-th/0110054} {arXiv:hep-th/0110054 [hep-th]}
  \BibitemShut {NoStop}%
\bibitem [{\citenamefont {Christiansen}\ \emph
  {et~al.}(2014{\natexlab{a}})\citenamefont {Christiansen}, \citenamefont
  {Litim}, \citenamefont {Pawlowski},\ and\ \citenamefont
  {Rodigast}}]{Christiansen:2012rx}%
  \BibitemOpen
  \bibfield  {author} {\bibinfo {author} {\bibfnamefont {N.}~\bibnamefont
  {Christiansen}}, \bibinfo {author} {\bibfnamefont {D.~F.}\ \bibnamefont
  {Litim}}, \bibinfo {author} {\bibfnamefont {J.~M.}\ \bibnamefont
  {Pawlowski}}, \ and\ \bibinfo {author} {\bibfnamefont {A.}~\bibnamefont
  {Rodigast}},\ }\href {\doibase 10.1016/j.physletb.2013.11.025} {\bibfield
  {journal} {\bibinfo  {journal} {Phys.Lett.}\ }\textbf {\bibinfo {volume}
  {B728}},\ \bibinfo {pages} {114} (\bibinfo {year} {2014}{\natexlab{a}})},\
  \Eprint {http://arxiv.org/abs/1209.4038} {arXiv:1209.4038 [hep-th]}
  \BibitemShut {NoStop}%
\bibitem [{\citenamefont {Christiansen}\ \emph
  {et~al.}(2014{\natexlab{b}})\citenamefont {Christiansen}, \citenamefont
  {Knorr}, \citenamefont {Pawlowski},\ and\ \citenamefont
  {Rodigast}}]{Christiansen:2014raa}%
  \BibitemOpen
  \bibfield  {author} {\bibinfo {author} {\bibfnamefont {N.}~\bibnamefont
  {Christiansen}}, \bibinfo {author} {\bibfnamefont {B.}~\bibnamefont {Knorr}},
  \bibinfo {author} {\bibfnamefont {J.~M.}\ \bibnamefont {Pawlowski}}, \ and\
  \bibinfo {author} {\bibfnamefont {A.}~\bibnamefont {Rodigast}},\ }\href@noop
  {} {\  (\bibinfo {year} {2014}{\natexlab{b}})},\ \Eprint
  {http://arxiv.org/abs/1403.1232} {arXiv:1403.1232 [hep-th]} \BibitemShut
  {NoStop}%
\bibitem [{\citenamefont {Lauscher}\ and\ \citenamefont
  {Reuter}(2002)}]{Lauscher:2002sq}%
  \BibitemOpen
  \bibfield  {author} {\bibinfo {author} {\bibfnamefont {O.}~\bibnamefont
  {Lauscher}}\ and\ \bibinfo {author} {\bibfnamefont {M.}~\bibnamefont
  {Reuter}},\ }\href {\doibase 10.1103/PhysRevD.66.025026} {\bibfield
  {journal} {\bibinfo  {journal} {Phys. Rev.}\ }\textbf {\bibinfo {volume}
  {D66}},\ \bibinfo {pages} {025026} (\bibinfo {year} {2002})},\ \Eprint
  {http://arxiv.org/abs/hep-th/0205062} {arXiv:hep-th/0205062} \BibitemShut
  {NoStop}%
\bibitem [{\citenamefont {Codello}\ and\ \citenamefont
  {Percacci}(2006)}]{Codello:2006in}%
  \BibitemOpen
  \bibfield  {author} {\bibinfo {author} {\bibfnamefont {A.}~\bibnamefont
  {Codello}}\ and\ \bibinfo {author} {\bibfnamefont {R.}~\bibnamefont
  {Percacci}},\ }\href {\doibase 10.1103/PhysRevLett.97.221301} {\bibfield
  {journal} {\bibinfo  {journal} {Phys. Rev. Lett.}\ }\textbf {\bibinfo
  {volume} {97}},\ \bibinfo {pages} {221301} (\bibinfo {year} {2006})},\
  \Eprint {http://arxiv.org/abs/hep-th/0607128} {arXiv:hep-th/0607128}
  \BibitemShut {NoStop}%
\bibitem [{\citenamefont {Codello}\ \emph {et~al.}(2008)\citenamefont
  {Codello}, \citenamefont {Percacci},\ and\ \citenamefont
  {Rahmede}}]{Codello:2007bd}%
  \BibitemOpen
  \bibfield  {author} {\bibinfo {author} {\bibfnamefont {A.}~\bibnamefont
  {Codello}}, \bibinfo {author} {\bibfnamefont {R.}~\bibnamefont {Percacci}}, \
  and\ \bibinfo {author} {\bibfnamefont {C.}~\bibnamefont {Rahmede}},\ }\href
  {\doibase 10.1142/S0217751X08038135} {\bibfield  {journal} {\bibinfo
  {journal} {Int. J. Mod. Phys.}\ }\textbf {\bibinfo {volume} {A23}},\ \bibinfo
  {pages} {143} (\bibinfo {year} {2008})},\ \Eprint
  {http://arxiv.org/abs/0705.1769} {arXiv:0705.1769 [hep-th]} \BibitemShut
  {NoStop}%
\bibitem [{\citenamefont {Codello}\ \emph {et~al.}(2009)\citenamefont
  {Codello}, \citenamefont {Percacci},\ and\ \citenamefont
  {Rahmede}}]{Codello:2008vh}%
  \BibitemOpen
  \bibfield  {author} {\bibinfo {author} {\bibfnamefont {A.}~\bibnamefont
  {Codello}}, \bibinfo {author} {\bibfnamefont {R.}~\bibnamefont {Percacci}}, \
  and\ \bibinfo {author} {\bibfnamefont {C.}~\bibnamefont {Rahmede}},\ }\href
  {\doibase 10.1016/j.aop.2008.08.008} {\bibfield  {journal} {\bibinfo
  {journal} {Annals Phys.}\ }\textbf {\bibinfo {volume} {324}},\ \bibinfo
  {pages} {414} (\bibinfo {year} {2009})},\ \Eprint
  {http://arxiv.org/abs/0805.2909} {arXiv:0805.2909 [hep-th]} \BibitemShut
  {NoStop}%
\bibitem [{\citenamefont {Machado}\ and\ \citenamefont
  {Saueressig}(2008)}]{Machado:2007ea}%
  \BibitemOpen
  \bibfield  {author} {\bibinfo {author} {\bibfnamefont {P.~F.}\ \bibnamefont
  {Machado}}\ and\ \bibinfo {author} {\bibfnamefont {F.}~\bibnamefont
  {Saueressig}},\ }\href {\doibase 10.1103/PhysRevD.77.124045} {\bibfield
  {journal} {\bibinfo  {journal} {Phys. Rev.}\ }\textbf {\bibinfo {volume}
  {D77}},\ \bibinfo {pages} {124045} (\bibinfo {year} {2008})},\ \Eprint
  {http://arxiv.org/abs/0712.0445} {arXiv:0712.0445 [hep-th]} \BibitemShut
  {NoStop}%
\bibitem [{\citenamefont {Benedetti}\ \emph {et~al.}(2009)\citenamefont
  {Benedetti}, \citenamefont {Machado},\ and\ \citenamefont
  {Saueressig}}]{Benedetti:2009rx}%
  \BibitemOpen
  \bibfield  {author} {\bibinfo {author} {\bibfnamefont {D.}~\bibnamefont
  {Benedetti}}, \bibinfo {author} {\bibfnamefont {P.~F.}\ \bibnamefont
  {Machado}}, \ and\ \bibinfo {author} {\bibfnamefont {F.}~\bibnamefont
  {Saueressig}},\ }\href {\doibase 10.1142/S0217732309031521} {\bibfield
  {journal} {\bibinfo  {journal} {Mod. Phys. Lett.}\ }\textbf {\bibinfo
  {volume} {A24}},\ \bibinfo {pages} {2233} (\bibinfo {year} {2009})},\ \Eprint
  {http://arxiv.org/abs/0901.2984} {arXiv:0901.2984 [hep-th]} \BibitemShut
  {NoStop}%
\bibitem [{\citenamefont {Eichhorn}\ \emph {et~al.}(2009)\citenamefont
  {Eichhorn}, \citenamefont {Gies},\ and\ \citenamefont
  {Scherer}}]{Eichhorn:2009ah}%
  \BibitemOpen
  \bibfield  {author} {\bibinfo {author} {\bibfnamefont {A.}~\bibnamefont
  {Eichhorn}}, \bibinfo {author} {\bibfnamefont {H.}~\bibnamefont {Gies}}, \
  and\ \bibinfo {author} {\bibfnamefont {M.~M.}\ \bibnamefont {Scherer}},\
  }\href {\doibase 10.1103/PhysRevD.80.104003} {\bibfield  {journal} {\bibinfo
  {journal} {Phys. Rev.}\ }\textbf {\bibinfo {volume} {D80}},\ \bibinfo {pages}
  {104003} (\bibinfo {year} {2009})},\ \Eprint {http://arxiv.org/abs/0907.1828}
  {arXiv:0907.1828 [hep-th]} \BibitemShut {NoStop}%
\bibitem [{\citenamefont {Manrique}\ \emph {et~al.}(2011)\citenamefont
  {Manrique}, \citenamefont {Rechenberger},\ and\ \citenamefont
  {Saueressig}}]{Manrique:2011jc}%
  \BibitemOpen
  \bibfield  {author} {\bibinfo {author} {\bibfnamefont {E.}~\bibnamefont
  {Manrique}}, \bibinfo {author} {\bibfnamefont {S.}~\bibnamefont
  {Rechenberger}}, \ and\ \bibinfo {author} {\bibfnamefont {F.}~\bibnamefont
  {Saueressig}},\ }\href {\doibase 10.1103/PhysRevLett.106.251302} {\bibfield
  {journal} {\bibinfo  {journal} {Phys.Rev.Lett.}\ }\textbf {\bibinfo {volume}
  {106}},\ \bibinfo {pages} {251302} (\bibinfo {year} {2011})},\ \Eprint
  {http://arxiv.org/abs/1102.5012} {arXiv:1102.5012 [hep-th]} \BibitemShut
  {NoStop}%
\bibitem [{\citenamefont {Rechenberger}\ and\ \citenamefont
  {Saueressig}(2012)}]{Rechenberger:2012pm}%
  \BibitemOpen
  \bibfield  {author} {\bibinfo {author} {\bibfnamefont {S.}~\bibnamefont
  {Rechenberger}}\ and\ \bibinfo {author} {\bibfnamefont {F.}~\bibnamefont
  {Saueressig}},\ }\href {\doibase 10.1103/PhysRevD.86.024018} {\bibfield
  {journal} {\bibinfo  {journal} {Phys.Rev.}\ }\textbf {\bibinfo {volume}
  {D86}},\ \bibinfo {pages} {024018} (\bibinfo {year} {2012})},\ \Eprint
  {http://arxiv.org/abs/1206.0657} {arXiv:1206.0657 [hep-th]} \BibitemShut
  {NoStop}%
\bibitem [{\citenamefont {Donkin}\ and\ \citenamefont
  {Pawlowski}(2012)}]{Donkin:2012ud}%
  \BibitemOpen
  \bibfield  {author} {\bibinfo {author} {\bibfnamefont {I.}~\bibnamefont
  {Donkin}}\ and\ \bibinfo {author} {\bibfnamefont {J.~M.}\ \bibnamefont
  {Pawlowski}},\ }\href@noop {} {\  (\bibinfo {year} {2012})},\ \Eprint
  {http://arxiv.org/abs/1203.4207} {arXiv:1203.4207 [hep-th]} \BibitemShut
  {NoStop}%
\bibitem [{\citenamefont {Codello}\ \emph {et~al.}(2014)\citenamefont
  {Codello}, \citenamefont {D'Odorico},\ and\ \citenamefont
  {Pagani}}]{Codello:2013fpa}%
  \BibitemOpen
  \bibfield  {author} {\bibinfo {author} {\bibfnamefont {A.}~\bibnamefont
  {Codello}}, \bibinfo {author} {\bibfnamefont {G.}~\bibnamefont {D'Odorico}},
  \ and\ \bibinfo {author} {\bibfnamefont {C.}~\bibnamefont {Pagani}},\ }\href
  {\doibase 10.1103/PhysRevD.89.081701} {\bibfield  {journal} {\bibinfo
  {journal} {Phys. Rev.}\ }\textbf {\bibinfo {volume} {D89}},\ \bibinfo {pages}
  {081701} (\bibinfo {year} {2014})},\ \Eprint {http://arxiv.org/abs/1304.4777}
  {arXiv:1304.4777 [gr-qc]} \BibitemShut {NoStop}%
\bibitem [{\citenamefont {Falls}\ \emph {et~al.}(2013)\citenamefont {Falls},
  \citenamefont {Litim}, \citenamefont {Nikolakopoulos},\ and\ \citenamefont
  {Rahmede}}]{Falls:2013bv}%
  \BibitemOpen
  \bibfield  {author} {\bibinfo {author} {\bibfnamefont {K.}~\bibnamefont
  {Falls}}, \bibinfo {author} {\bibfnamefont {D.}~\bibnamefont {Litim}},
  \bibinfo {author} {\bibfnamefont {K.}~\bibnamefont {Nikolakopoulos}}, \ and\
  \bibinfo {author} {\bibfnamefont {C.}~\bibnamefont {Rahmede}},\ }\href@noop
  {} {\  (\bibinfo {year} {2013})},\ \Eprint {http://arxiv.org/abs/1301.4191}
  {arXiv:1301.4191 [hep-th]} \BibitemShut {NoStop}%
\bibitem [{\citenamefont {Falls}(2014)}]{Falls:2014zba}%
  \BibitemOpen
  \bibfield  {author} {\bibinfo {author} {\bibfnamefont {K.}~\bibnamefont
  {Falls}},\ }\href@noop {} {\  (\bibinfo {year} {2014})},\ \Eprint
  {http://arxiv.org/abs/1408.0276} {arXiv:1408.0276 [hep-th]} \BibitemShut
  {NoStop}%
\bibitem [{\citenamefont {Falls}\ \emph {et~al.}(2014)\citenamefont {Falls},
  \citenamefont {Litim}, \citenamefont {Nikolakopoulos},\ and\ \citenamefont
  {Rahmede}}]{Falls:2014tra}%
  \BibitemOpen
  \bibfield  {author} {\bibinfo {author} {\bibfnamefont {K.}~\bibnamefont
  {Falls}}, \bibinfo {author} {\bibfnamefont {D.~F.}\ \bibnamefont {Litim}},
  \bibinfo {author} {\bibfnamefont {K.}~\bibnamefont {Nikolakopoulos}}, \ and\
  \bibinfo {author} {\bibfnamefont {C.}~\bibnamefont {Rahmede}},\ }\href@noop
  {} {\  (\bibinfo {year} {2014})},\ \Eprint {http://arxiv.org/abs/1410.4815}
  {arXiv:1410.4815 [hep-th]} \BibitemShut {NoStop}%
\bibitem [{\citenamefont {Falls}(2015)}]{Falls:2015qga}%
  \BibitemOpen
  \bibfield  {author} {\bibinfo {author} {\bibfnamefont {K.}~\bibnamefont
  {Falls}},\ }\href@noop {} {\  (\bibinfo {year} {2015})},\ \Eprint
  {http://arxiv.org/abs/1501.05331} {arXiv:1501.05331 [hep-th]} \BibitemShut
  {NoStop}%
\bibitem [{\citenamefont {Gies}\ \emph {et~al.}(2015)\citenamefont {Gies},
  \citenamefont {Knorr},\ and\ \citenamefont {Lippoldt}}]{Gies:2015tca}%
  \BibitemOpen
  \bibfield  {author} {\bibinfo {author} {\bibfnamefont {H.}~\bibnamefont
  {Gies}}, \bibinfo {author} {\bibfnamefont {B.}~\bibnamefont {Knorr}}, \ and\
  \bibinfo {author} {\bibfnamefont {S.}~\bibnamefont {Lippoldt}},\ }\href
  {\doibase 10.1103/PhysRevD.92.084020} {\bibfield  {journal} {\bibinfo
  {journal} {Phys. Rev.}\ }\textbf {\bibinfo {volume} {D92}},\ \bibinfo {pages}
  {084020} (\bibinfo {year} {2015})},\ \Eprint
  {http://arxiv.org/abs/1507.08859} {arXiv:1507.08859 [hep-th]} \BibitemShut
  {NoStop}%
\bibitem [{\citenamefont {Gies}\ \emph {et~al.}(2016)\citenamefont {Gies},
  \citenamefont {Knorr}, \citenamefont {Lippoldt},\ and\ \citenamefont
  {Saueressig}}]{Gies:2016con}%
  \BibitemOpen
  \bibfield  {author} {\bibinfo {author} {\bibfnamefont {H.}~\bibnamefont
  {Gies}}, \bibinfo {author} {\bibfnamefont {B.}~\bibnamefont {Knorr}},
  \bibinfo {author} {\bibfnamefont {S.}~\bibnamefont {Lippoldt}}, \ and\
  \bibinfo {author} {\bibfnamefont {F.}~\bibnamefont {Saueressig}},\
  }\href@noop {} {\  (\bibinfo {year} {2016})},\ \Eprint
  {http://arxiv.org/abs/1601.01800} {arXiv:1601.01800 [hep-th]} \BibitemShut
  {NoStop}%
\bibitem [{\citenamefont {Niedermaier}\ and\ \citenamefont
  {Reuter}(2006)}]{Niedermaier:2006wt}%
  \BibitemOpen
  \bibfield  {author} {\bibinfo {author} {\bibfnamefont {M.}~\bibnamefont
  {Niedermaier}}\ and\ \bibinfo {author} {\bibfnamefont {M.}~\bibnamefont
  {Reuter}},\ }\href@noop {} {\bibfield  {journal} {\bibinfo  {journal} {Living
  Rev.Rel.}\ }\textbf {\bibinfo {volume} {9}},\ \bibinfo {pages} {5} (\bibinfo
  {year} {2006})}\BibitemShut {NoStop}%
\bibitem [{\citenamefont {Percacci}(2007)}]{Percacci:2007sz}%
  \BibitemOpen
  \bibfield  {author} {\bibinfo {author} {\bibfnamefont {R.}~\bibnamefont
  {Percacci}},\ }\href@noop {} {\bibfield  {journal} {\bibinfo  {journal} {In
  *Oriti, D. (ed.): Approaches to quantum gravity* 111-128}\ } (\bibinfo {year}
  {2007})},\ \Eprint {http://arxiv.org/abs/0709.3851} {arXiv:0709.3851
  [hep-th]} \BibitemShut {NoStop}%
\bibitem [{\citenamefont {Litim}(2011)}]{Litim:2011cp}%
  \BibitemOpen
  \bibfield  {author} {\bibinfo {author} {\bibfnamefont {D.~F.}\ \bibnamefont
  {Litim}},\ }\href@noop {} {\bibfield  {journal} {\bibinfo  {journal}
  {Phil.Trans.Roy.Soc.Lond.}\ }\textbf {\bibinfo {volume} {A369}},\ \bibinfo
  {pages} {2759} (\bibinfo {year} {2011})},\ \Eprint
  {http://arxiv.org/abs/1102.4624} {arXiv:1102.4624 [hep-th]} \BibitemShut
  {NoStop}%
\bibitem [{\citenamefont {Reuter}\ and\ \citenamefont
  {Saueressig}(2012)}]{Reuter:2012id}%
  \BibitemOpen
  \bibfield  {author} {\bibinfo {author} {\bibfnamefont {M.}~\bibnamefont
  {Reuter}}\ and\ \bibinfo {author} {\bibfnamefont {F.}~\bibnamefont
  {Saueressig}},\ }\href {\doibase 10.1088/1367-2630/14/5/055022} {\bibfield
  {journal} {\bibinfo  {journal} {New J. Phys.}\ }\textbf {\bibinfo {volume}
  {14}},\ \bibinfo {pages} {055022} (\bibinfo {year} {2012})},\ \Eprint
  {http://arxiv.org/abs/1202.2274} {arXiv:1202.2274 [hep-th]} \BibitemShut
  {NoStop}%
\bibitem [{\citenamefont {Dou}\ and\ \citenamefont
  {Percacci}(1998)}]{Dou:1997fg}%
  \BibitemOpen
  \bibfield  {author} {\bibinfo {author} {\bibfnamefont {D.}~\bibnamefont
  {Dou}}\ and\ \bibinfo {author} {\bibfnamefont {R.}~\bibnamefont {Percacci}},\
  }\href {\doibase 10.1088/0264-9381/15/11/011} {\bibfield  {journal} {\bibinfo
   {journal} {Class. Quant. Grav.}\ }\textbf {\bibinfo {volume} {15}},\
  \bibinfo {pages} {3449} (\bibinfo {year} {1998})},\ \Eprint
  {http://arxiv.org/abs/hep-th/9707239} {arXiv:hep-th/9707239 [hep-th]}
  \BibitemShut {NoStop}%
\bibitem [{\citenamefont {Percacci}\ and\ \citenamefont
  {Perini}(2003{\natexlab{a}})}]{Percacci:2002ie}%
  \BibitemOpen
  \bibfield  {author} {\bibinfo {author} {\bibfnamefont {R.}~\bibnamefont
  {Percacci}}\ and\ \bibinfo {author} {\bibfnamefont {D.}~\bibnamefont
  {Perini}},\ }\href {\doibase 10.1103/PhysRevD.67.081503} {\bibfield
  {journal} {\bibinfo  {journal} {Phys. Rev.}\ }\textbf {\bibinfo {volume}
  {D67}},\ \bibinfo {pages} {081503} (\bibinfo {year} {2003}{\natexlab{a}})},\
  \Eprint {http://arxiv.org/abs/hep-th/0207033} {arXiv:hep-th/0207033}
  \BibitemShut {NoStop}%
\bibitem [{\citenamefont {Percacci}\ and\ \citenamefont
  {Perini}(2003{\natexlab{b}})}]{Percacci:2003jz}%
  \BibitemOpen
  \bibfield  {author} {\bibinfo {author} {\bibfnamefont {R.}~\bibnamefont
  {Percacci}}\ and\ \bibinfo {author} {\bibfnamefont {D.}~\bibnamefont
  {Perini}},\ }\href {\doibase 10.1103/PhysRevD.68.044018} {\bibfield
  {journal} {\bibinfo  {journal} {Phys. Rev.}\ }\textbf {\bibinfo {volume}
  {D68}},\ \bibinfo {pages} {044018} (\bibinfo {year} {2003}{\natexlab{b}})},\
  \Eprint {http://arxiv.org/abs/hep-th/0304222} {arXiv:hep-th/0304222}
  \BibitemShut {NoStop}%
\bibitem [{\citenamefont {Folkerts}\ \emph {et~al.}(2012)\citenamefont
  {Folkerts}, \citenamefont {Litim},\ and\ \citenamefont
  {Pawlowski}}]{Folkerts:2011jz}%
  \BibitemOpen
  \bibfield  {author} {\bibinfo {author} {\bibfnamefont {S.}~\bibnamefont
  {Folkerts}}, \bibinfo {author} {\bibfnamefont {D.~F.}\ \bibnamefont {Litim}},
  \ and\ \bibinfo {author} {\bibfnamefont {J.~M.}\ \bibnamefont {Pawlowski}},\
  }\href {\doibase 10.1016/j.physletb.2012.02.002} {\bibfield  {journal}
  {\bibinfo  {journal} {Phys.Lett.}\ }\textbf {\bibinfo {volume} {B709}},\
  \bibinfo {pages} {234} (\bibinfo {year} {2012})},\ \Eprint
  {http://arxiv.org/abs/1101.5552} {arXiv:1101.5552 [hep-th]} \BibitemShut
  {NoStop}%
\bibitem [{\citenamefont {Don{\`a}}\ and\ \citenamefont
  {Percacci}(2013)}]{Dona:2012am}%
  \BibitemOpen
  \bibfield  {author} {\bibinfo {author} {\bibfnamefont {P.}~\bibnamefont
  {Don{\`a}}}\ and\ \bibinfo {author} {\bibfnamefont {R.}~\bibnamefont
  {Percacci}},\ }\href {\doibase 10.1103/PhysRevD.87.045002} {\bibfield
  {journal} {\bibinfo  {journal} {Phys. Rev.}\ }\textbf {\bibinfo {volume}
  {D87}},\ \bibinfo {pages} {045002} (\bibinfo {year} {2013})},\ \Eprint
  {http://arxiv.org/abs/1209.3649} {arXiv:1209.3649 [hep-th]} \BibitemShut
  {NoStop}%
\bibitem [{\citenamefont {Don{\`a}}\ \emph {et~al.}(2014)\citenamefont
  {Don{\`a}}, \citenamefont {Eichhorn},\ and\ \citenamefont
  {Percacci}}]{Dona:2013qba}%
  \BibitemOpen
  \bibfield  {author} {\bibinfo {author} {\bibfnamefont {P.}~\bibnamefont
  {Don{\`a}}}, \bibinfo {author} {\bibfnamefont {A.}~\bibnamefont {Eichhorn}},
  \ and\ \bibinfo {author} {\bibfnamefont {R.}~\bibnamefont {Percacci}},\
  }\href {\doibase 10.1103/PhysRevD.89.084035} {\bibfield  {journal} {\bibinfo
  {journal} {Phys.Rev.}\ }\textbf {\bibinfo {volume} {D89}},\ \bibinfo {pages}
  {084035} (\bibinfo {year} {2014})},\ \Eprint {http://arxiv.org/abs/1311.2898}
  {arXiv:1311.2898 [hep-th]} \BibitemShut {NoStop}%
\bibitem [{\citenamefont {Donà}\ \emph
  {et~al.}(2015{\natexlab{a}})\citenamefont {Donà}, \citenamefont {Eichhorn},\
  and\ \citenamefont {Percacci}}]{Dona:2014pla}%
  \BibitemOpen
  \bibfield  {author} {\bibinfo {author} {\bibfnamefont {P.}~\bibnamefont
  {Donà}}, \bibinfo {author} {\bibfnamefont {A.}~\bibnamefont {Eichhorn}}, \
  and\ \bibinfo {author} {\bibfnamefont {R.}~\bibnamefont {Percacci}},\
  }\bibfield  {booktitle} {\emph {\bibinfo {booktitle} {{Proceedings, Satellite
  Conference on Theory Canada 9}}},\ }\href {\doibase 10.1139/cjp-2014-0574}
  {\bibfield  {journal} {\bibinfo  {journal} {Can. J. Phys.}\ }\textbf
  {\bibinfo {volume} {93}},\ \bibinfo {pages} {988} (\bibinfo {year}
  {2015}{\natexlab{a}})},\ \Eprint {http://arxiv.org/abs/1410.4411}
  {arXiv:1410.4411 [gr-qc]} \BibitemShut {NoStop}%
\bibitem [{\citenamefont {Oda}\ and\ \citenamefont
  {Yamada}(2015)}]{Oda:2015sma}%
  \BibitemOpen
  \bibfield  {author} {\bibinfo {author} {\bibfnamefont {K.-y.}\ \bibnamefont
  {Oda}}\ and\ \bibinfo {author} {\bibfnamefont {M.}~\bibnamefont {Yamada}},\
  }\href@noop {} {\  (\bibinfo {year} {2015})},\ \Eprint
  {http://arxiv.org/abs/1510.03734} {arXiv:1510.03734 [hep-th]} \BibitemShut
  {NoStop}%
\bibitem [{\citenamefont {Donà}\ \emph
  {et~al.}(2015{\natexlab{b}})\citenamefont {Donà}, \citenamefont {Eichhorn},
  \citenamefont {Labus},\ and\ \citenamefont {Percacci}}]{Dona:2015tnf}%
  \BibitemOpen
  \bibfield  {author} {\bibinfo {author} {\bibfnamefont {P.}~\bibnamefont
  {Donà}}, \bibinfo {author} {\bibfnamefont {A.}~\bibnamefont {Eichhorn}},
  \bibinfo {author} {\bibfnamefont {P.}~\bibnamefont {Labus}}, \ and\ \bibinfo
  {author} {\bibfnamefont {R.}~\bibnamefont {Percacci}},\ }\href@noop {} {\
  (\bibinfo {year} {2015}{\natexlab{b}})},\ \Eprint
  {http://arxiv.org/abs/1512.01589} {arXiv:1512.01589 [gr-qc]} \BibitemShut
  {NoStop}%
\bibitem [{\citenamefont {Eichhorn}\ and\ \citenamefont
  {Gies}(2011)}]{Eichhorn:2011pc}%
  \BibitemOpen
  \bibfield  {author} {\bibinfo {author} {\bibfnamefont {A.}~\bibnamefont
  {Eichhorn}}\ and\ \bibinfo {author} {\bibfnamefont {H.}~\bibnamefont
  {Gies}},\ }\href {\doibase 10.1088/1367-2630/13/12/125012} {\bibfield
  {journal} {\bibinfo  {journal} {New J. Phys.}\ }\textbf {\bibinfo {volume}
  {13}},\ \bibinfo {pages} {125012} (\bibinfo {year} {2011})},\ \Eprint
  {http://arxiv.org/abs/1104.5366} {arXiv:1104.5366 [hep-th]} \BibitemShut
  {NoStop}%
\bibitem [{\citenamefont {Eichhorn}(2012)}]{Eichhorn:2012va}%
  \BibitemOpen
  \bibfield  {author} {\bibinfo {author} {\bibfnamefont {A.}~\bibnamefont
  {Eichhorn}},\ }\href {\doibase 10.1103/PhysRevD.86.105021} {\bibfield
  {journal} {\bibinfo  {journal} {Phys. Rev.}\ }\textbf {\bibinfo {volume}
  {D86}},\ \bibinfo {pages} {105021} (\bibinfo {year} {2012})},\ \Eprint
  {http://arxiv.org/abs/1204.0965} {arXiv:1204.0965 [gr-qc]} \BibitemShut
  {NoStop}%
\bibitem [{\citenamefont {Henz}\ \emph {et~al.}(2013)\citenamefont {Henz},
  \citenamefont {Pawlowski}, \citenamefont {Rodigast},\ and\ \citenamefont
  {Wetterich}}]{Henz:2013oxa}%
  \BibitemOpen
  \bibfield  {author} {\bibinfo {author} {\bibfnamefont {T.}~\bibnamefont
  {Henz}}, \bibinfo {author} {\bibfnamefont {J.~M.}\ \bibnamefont {Pawlowski}},
  \bibinfo {author} {\bibfnamefont {A.}~\bibnamefont {Rodigast}}, \ and\
  \bibinfo {author} {\bibfnamefont {C.}~\bibnamefont {Wetterich}},\ }\href
  {\doibase 10.1016/j.physletb.2013.10.015} {\bibfield  {journal} {\bibinfo
  {journal} {Phys. Lett.}\ }\textbf {\bibinfo {volume} {B727}},\ \bibinfo
  {pages} {298} (\bibinfo {year} {2013})},\ \Eprint
  {http://arxiv.org/abs/1304.7743} {arXiv:1304.7743 [hep-th]} \BibitemShut
  {NoStop}%
\bibitem [{\citenamefont {Wetterich}(1993)}]{Wetterich:1992yh}%
  \BibitemOpen
  \bibfield  {author} {\bibinfo {author} {\bibfnamefont {C.}~\bibnamefont
  {Wetterich}},\ }\href {\doibase 10.1016/0370-2693(93)90726-X} {\bibfield
  {journal} {\bibinfo  {journal} {Phys.Lett.}\ }\textbf {\bibinfo {volume}
  {B301}},\ \bibinfo {pages} {90} (\bibinfo {year} {1993})}\BibitemShut
  {NoStop}%
\bibitem [{\citenamefont {Litim}\ and\ \citenamefont
  {Pawlowski}(1998)}]{Litim:1998qi}%
  \BibitemOpen
  \bibfield  {author} {\bibinfo {author} {\bibfnamefont {D.~F.}\ \bibnamefont
  {Litim}}\ and\ \bibinfo {author} {\bibfnamefont {J.~M.}\ \bibnamefont
  {Pawlowski}},\ }\href {\doibase 10.1016/S0370-2693(98)00761-8} {\bibfield
  {journal} {\bibinfo  {journal} {Phys.Lett.}\ }\textbf {\bibinfo {volume}
  {B435}},\ \bibinfo {pages} {181} (\bibinfo {year} {1998})},\ \Eprint
  {http://arxiv.org/abs/hep-th/9802064} {arXiv:hep-th/9802064 [hep-th]}
  \BibitemShut {NoStop}%
\bibitem [{\citenamefont {Weldon}(2001)}]{Weldon:2000fr}%
  \BibitemOpen
  \bibfield  {author} {\bibinfo {author} {\bibfnamefont {H.~A.}\ \bibnamefont
  {Weldon}},\ }\href {\doibase 10.1103/PhysRevD.63.104010} {\bibfield
  {journal} {\bibinfo  {journal} {Phys. Rev.}\ }\textbf {\bibinfo {volume}
  {D63}},\ \bibinfo {pages} {104010} (\bibinfo {year} {2001})},\ \Eprint
  {http://arxiv.org/abs/gr-qc/0009086} {arXiv:gr-qc/0009086 [gr-qc]}
  \BibitemShut {NoStop}%
\bibitem [{\citenamefont {Gies}\ and\ \citenamefont
  {Lippoldt}(2014)}]{Gies:2013noa}%
  \BibitemOpen
  \bibfield  {author} {\bibinfo {author} {\bibfnamefont {H.}~\bibnamefont
  {Gies}}\ and\ \bibinfo {author} {\bibfnamefont {S.}~\bibnamefont
  {Lippoldt}},\ }\href {\doibase 10.1103/PhysRevD.89.064040} {\bibfield
  {journal} {\bibinfo  {journal} {Phys.Rev.}\ }\textbf {\bibinfo {volume}
  {D89}},\ \bibinfo {pages} {064040} (\bibinfo {year} {2014})},\ \Eprint
  {http://arxiv.org/abs/1310.2509} {arXiv:1310.2509 [hep-th]} \BibitemShut
  {NoStop}%
\bibitem [{\citenamefont {Lippoldt}(2015)}]{Lippoldt:2015cea}%
  \BibitemOpen
  \bibfield  {author} {\bibinfo {author} {\bibfnamefont {S.}~\bibnamefont
  {Lippoldt}},\ }\href {\doibase 10.1103/PhysRevD.91.104006} {\bibfield
  {journal} {\bibinfo  {journal} {Phys. Rev.}\ }\textbf {\bibinfo {volume}
  {D91}},\ \bibinfo {pages} {104006} (\bibinfo {year} {2015})},\ \Eprint
  {http://arxiv.org/abs/1502.05607} {arXiv:1502.05607 [hep-th]} \BibitemShut
  {NoStop}%
\bibitem [{\citenamefont {Mitter}\ \emph {et~al.}(2015)\citenamefont {Mitter},
  \citenamefont {Pawlowski},\ and\ \citenamefont
  {Strodthoff}}]{Mitter:2014wpa}%
  \BibitemOpen
  \bibfield  {author} {\bibinfo {author} {\bibfnamefont {M.}~\bibnamefont
  {Mitter}}, \bibinfo {author} {\bibfnamefont {J.~M.}\ \bibnamefont
  {Pawlowski}}, \ and\ \bibinfo {author} {\bibfnamefont {N.}~\bibnamefont
  {Strodthoff}},\ }\href {\doibase 10.1103/PhysRevD.91.054035} {\bibfield
  {journal} {\bibinfo  {journal} {Phys. Rev.}\ }\textbf {\bibinfo {volume}
  {D91}},\ \bibinfo {pages} {054035} (\bibinfo {year} {2015})},\ \Eprint
  {http://arxiv.org/abs/1411.7978} {arXiv:1411.7978 [hep-ph]} \BibitemShut
  {NoStop}%
\bibitem [{\citenamefont {Gies}\ and\ \citenamefont
  {Jaeckel}(2006)}]{Gies:2005as}%
  \BibitemOpen
  \bibfield  {author} {\bibinfo {author} {\bibfnamefont {H.}~\bibnamefont
  {Gies}}\ and\ \bibinfo {author} {\bibfnamefont {J.}~\bibnamefont {Jaeckel}},\
  }\href {\doibase 10.1140/epjc/s2006-02475-0} {\bibfield  {journal} {\bibinfo
  {journal} {Eur. Phys. J.}\ }\textbf {\bibinfo {volume} {C46}},\ \bibinfo
  {pages} {433} (\bibinfo {year} {2006})},\ \Eprint
  {http://arxiv.org/abs/hep-ph/0507171} {arXiv:hep-ph/0507171 [hep-ph]}
  \BibitemShut {NoStop}%
\bibitem [{\citenamefont {Braun}(2012)}]{Braun:2011pp}%
  \BibitemOpen
  \bibfield  {author} {\bibinfo {author} {\bibfnamefont {J.}~\bibnamefont
  {Braun}},\ }\href {\doibase 10.1088/0954-3899/39/3/033001} {\bibfield
  {journal} {\bibinfo  {journal} {J.Phys.}\ }\textbf {\bibinfo {volume}
  {G39}},\ \bibinfo {pages} {033001} (\bibinfo {year} {2012})},\ \Eprint
  {http://arxiv.org/abs/1108.4449} {arXiv:1108.4449 [hep-ph]} \BibitemShut
  {NoStop}%
\bibitem [{\citenamefont {Litim}(2000)}]{Litim:2000ci}%
  \BibitemOpen
  \bibfield  {author} {\bibinfo {author} {\bibfnamefont {D.~F.}\ \bibnamefont
  {Litim}},\ }\href {\doibase 10.1016/S0370-2693(00)00748-6} {\bibfield
  {journal} {\bibinfo  {journal} {Phys.Lett.}\ }\textbf {\bibinfo {volume}
  {B486}},\ \bibinfo {pages} {92} (\bibinfo {year} {2000})},\ \Eprint
  {http://arxiv.org/abs/hep-th/0005245} {arXiv:hep-th/0005245 [hep-th]}
  \BibitemShut {NoStop}%
\bibitem [{\citenamefont {Gies}\ \emph {et~al.}(2004)\citenamefont {Gies},
  \citenamefont {Jaeckel},\ and\ \citenamefont {Wetterich}}]{Gies:2003dp}%
  \BibitemOpen
  \bibfield  {author} {\bibinfo {author} {\bibfnamefont {H.}~\bibnamefont
  {Gies}}, \bibinfo {author} {\bibfnamefont {J.}~\bibnamefont {Jaeckel}}, \
  and\ \bibinfo {author} {\bibfnamefont {C.}~\bibnamefont {Wetterich}},\ }\href
  {\doibase 10.1103/PhysRevD.69.105008} {\bibfield  {journal} {\bibinfo
  {journal} {Phys. Rev.}\ }\textbf {\bibinfo {volume} {D69}},\ \bibinfo {pages}
  {105008} (\bibinfo {year} {2004})},\ \Eprint
  {http://arxiv.org/abs/hep-ph/0312034} {arXiv:hep-ph/0312034 [hep-ph]}
  \BibitemShut {NoStop}%
\end{thebibliography}
%
%
%
\end{document}